\documentstyle[preprint,eqsecnum,aps,tighten]{revtex}
\begin{document}
\draft
\preprint{NT@UW-99-21}

\title{%prelim. DRAFT lc/longaper.tex working title\\
Light-Front Nuclear Physics: Mean Field Theory for 
Finite Nuclei} 

\author{P.G.~Blunden}
\address{Department of Physics and Astronomy\\University of Manitoba\\
Winnipeg, MB, R3T~2N2\\Canada}

\author{M.~Burkardt}
\address{Department of Physics\\
New Mexico State University\\
Las Cruces, NM 88003-0001\\U.S.A.}

\author{G.A.~Miller}
\address{Department of Physics, Box 351560\\
University of Washington \\
Seattle, WA 98195-1560\\U.S.A.}

\date{\today}
\maketitle

\begin{abstract}
 
A light-front treatment for finite nuclei is developed from a
relativistic effective Lagrangian (QHD1) involving nucleons, scalar
mesons and vector mesons. We show that the necessary variational
principle is a constrained one which fixes the expectation value of the
total momentum operator $P^+$ to be the same as that for $P^-$. This is
the same as minimizing the sum of the total momentum operators:
$P^-+P^+$. We obtain a new light-front version of the equation that
defines the single nucleon modes. The solutions of this equation are
approximately a non-trivial phase factor times certain solutions of the
usual equal-time Dirac equation. The ground state wave function is
treated as a meson-nucleon Fock state, and the meson fields are treated
as expectation values of field operators in that ground state. The
resulting equations for these expectation values are shown to be
closely related to the usual meson field equations. A new numerical
technique to solve the self-consistent field equations is introduced
and applied to $^{16}$O and $^{40}$Ca. The computed binding energies
are essentially the same as for the usual equal-time theory. The
nucleon plus momentum distribution (probability for a nucleon to have a
given value of $p^+$) is obtained, and peaks for values of $p^+$ about
seventy percent of the nucleon mass. The mesonic component of the
ground state wave function is used to determine the scalar and vector
meson momentum distribution functions, with a result that the vector
mesons carry about thirty percent of the nuclear plus-momentum.
The vector meson momentum distribution becomes more concentrated at
$p^+=0$ as $A$ increases.

\end{abstract}

%\pacs{Valid PACS appear here}

\narrowtext

\section{Introduction}

The purpose of this paper is to derive the light-front formalism
necessary to compute the properties of finite nuclei. Nuclear
properties are very well handled within the existing conventional
nuclear theory, so it behooves us to explain why we are embarking on
this project. Our motivation is that understanding experiments
involving high energy nuclear reactions seems to require that
light-front dynamics and light cone variables be used. Consider the 
EMC experiment \cite{emc}, which showed that there is a significant
difference between the parton distributions of free nucleons and
nucleons in a nucleus. This difference can interpreted as a shift in 
the momentum distribution of valence quarks towards smaller values of
the Bjorken variable $x_{\text{Bj}}$. The Bjorken variable is a ratio
of the plus-momentum $k^+=k^0+k^3$ of a quark to that of the target.
Thus light cone variables are relevant. If one uses $k^+$ as a momentum
variable the corresponding canonical spatial variable is $x^-=x^0-x^3$
and the time variable is $x^+=x^0 +x^3$.

It is important to realize that the use of light-front dynamics is not
limited to quarks within the nucleon --- it also applies to nucleons
within the nucleus. This formalism is useful whenever the momentum of
initial or final state nucleons is large compared to their
mass \cite{fs}. In particular, it can be used for $(e,e'p)$ and $(p,2p)$
reactions. If one uses light-front variables for nucleons in a nucleus,
it is also necessary to maintain consistency with the information
derived previously using conventional nuclear dynamics. This provides
the technical challenge which we address in the present manuscript.

The conventional equal-time approach to nuclear structure physics
provides an excellent framework, so it is worthwhile to introduce the
light-front variables and describe the expected advantages in a general
way. The use of the light-cone variables can be obtained using a simple
argument based on kinematics \cite{fs}. Suppose the virtual photon that
is absorbed by a fermion at a space-time point ($z_1$, $t_1$). The
fermion then starts to move at high momentum and nearly the speed of
light and emits the photon at another space time point ($z_2$, $t_2$).
In between the two times, the wave function of the entire system has
undergone a time evolution given by the complicated operator
$e^{-iH(t_2-t_1)}$. But we have $z_1+ct_1=z_2+ct_2$, if the $z-$axis is
opposite to the direction of the virtual photon. The two scattering
events occur at different times, but at the same value of $x^+=z+ct$.
Thus if we use $x^+$ as a time variable, no time evolution factor
appears. The net result is that the cross section involves light-like
correlation functions which involve field operators evaluated at the
same light-front time: $x^+=0$ (see for example the reviews
 \cite{mb:adv,emcrevs}). Thus it is a specific and general feature of
the light-front wave approach that knowing only the ground state wave
function is sufficient for computing the distribution functions.

Let us review the salient features of the basic idea that using the
light-front approach leads to a simplified treatment. To be specific,
consider high energy electron scattering from nucleons in nuclei. The
key ingredient in the light-front simplification is to realize the main
difference between the two formalisms. In the equal-time formalism,
sums over intermediate states are taken over eigenstates of the
Hamiltonian, $P^0$. The usual three momentum is conserved, but energy
is not conserved in intermediate states. In the light-front approach
one sums over eigenstates of the minus component of the total momentum
operator. The value of $P^-$ is not conserved in intermediate state
sums, and the values of $P^+$ and $P^\perp$ are conserved. This is
especially convenient for high energy reactions, in which the
plus-component is the largest component of the momentum for each
projectile or ejectile.

The advantage of using $P^-$ as an ``energy'' variable can be easily
described. Let the four-momentum $q$ of the exchanged virtual photon be
given by $\left(\nu,0,0,-\sqrt{Q^2+\nu^2}\right)$, with $Q^2=-q^2$, and
$Q^2$ and $\nu^2$ are both very large but $Q^2/\nu$ is finite (the
Bjorken limit). In this case it is worthwhile to use the light-cone
variables $q^\pm=q^0\pm q^3$ in which $q^+\approx Q^2/2\nu=Mx$,
$q^-\approx 2\nu -Q^2/2\nu$, so that $q^-\gg q^+$. Here $M$ is the mass
of a nucleon and $x$ is the Bjorken variable. We shall neglect $q^-$ in
comparison to $q^+$, noting that corrections to this can be handled in
a systematic fashion. Then the schematic form of the scattering cross
section for $e+A\to e' +(A-1)_f +p$, where $f$ represents the final
nuclear eigenstate of $P^-$, and $p$ the four-momentum of the final
proton, is given by
\begin{equation}
d\sigma\sim \sum_f \int
{d^3p_f\over E_f}\int d^4p\, \delta (p^2-M^2)\delta^{(4)}(q+p_i-p_f-p)|
\langle p,f\mid J(q)\mid i\rangle\mid^2.
\end{equation}
Here the operator $J(q)$ is a schematic representation of the
electromagnetic current. Performing the four-dimensional integral over
$p$ leads to the expression
\begin{equation}
d\sigma\sim \sum_f \int {d^2p_fdp_f^+\over p^+_f}
\delta\left((p_i-p_f+q)^2-m^2\right)\mid \langle p,f\mid J(q)\mid
i\rangle\mid^2 \label{int}.
\end{equation}
The argument of the delta function $(p_i-p_f+q)^2-M^2 \approx
-Q^2+2q^-(p_i-p_f)^+$. Thus we see that $p_f^-$ does not appear in the
argument of the delta function, or anywhere else, so that we can
replace the sum over intermediate states by unity. In the usual
equal-time representation, one finds the argument of the delta function
to be $-Q^2+2\nu(E_i-E_f)$. The energy of the final state appears, and
one can not do the sum.

To proceed further in this schematic approach we take
\begin{equation}
J(q)=\int d^3k\, b^\dagger_{\bbox{k+q}} b_{\bbox{k}},\label{jq}
\end{equation}
where $b$ is a nucleon destruction operator and $\bbox{V}\equiv
(V^+,\bbox{V}_\perp)$. It is useful to define $\bbox{p_B}\equiv
\bbox{p_i}-\bbox{p_f}$ because
\begin{equation}
p_B^+=Q^2/2\nu\equiv M x ,\label{one}
\end{equation}
as demanded by the delta function. Then one can re-express
Eq.~(\ref{int}) as
\begin{equation}
d\sigma\sim d^2{p_B}_\perp \langle i\mid b^\dagger_{\bbox{p_B}}
\;b_{\bbox{p_B}}\mid i\rangle 
= d^2{p_B}_\perp n(M x,{p_B}_\perp),
\label{ds2}
\end{equation}
where $ n(M x,{p_B}_\perp)$ is the probability for a nucleon in
the ground state to have a momentum $(M x,{p_B}_\perp)$. Integration in
Eq.~(\ref{ds2}) leads to
\begin{equation}
\sigma \sim\int d^2p_\perp\, n(M x,{p}_\perp)\equiv f(Mx),
\end{equation}
with $f(Mx)$ as the probability for a nucleon in the ground state to
have a plus momentum of $Mx$. 

The quantity $f(Mx)$ has been a widely used prescription UP for
handling the light-front in a simple way. The variable $Mx$ is replaced
by $M-\varepsilon_\alpha +k^3$, in which the label $\alpha$ denotes a
shell-model orbital $\phi_\alpha$ of binding energy
$\varepsilon_\alpha$. Then
\begin{equation}
f_{UP}=\sum_{\alpha}n_\alpha\int d^2p_\perp\int dp^3\,\mid
\phi_\alpha(p^3,p_\perp)\mid^2 \delta(M-\varepsilon_\alpha +p^3-Mx),
\label{UP}
\end{equation}
in which $n_\alpha$ is an occupation probability. The validity of this
prescription, which rests on a reasonable assumption, is rather suspect
because the variable $p^+=Mx$ is a kinematic variable, unrelated to
discrete eigenvalues of a wave equation. One of the main purposes of
the present paper is to see if anything like this prescription emerges
from our calculations. We shall see that Eq.~(\ref{UP}) is not
obtained, if a vector potential is a significant part of the nuclear
mean field.

It is useful to discuss the relation with $y$-scaling \cite{west}. The
arguments that the cross section depends on a plus-momentum
distribution are well known when used for quarks in a nucleon, but they
also apply to nucleons in a nucleus \cite{fs}. Ji and Filippone \cite{jf}
showed that the $y$ scaling function $F(y)$ extracted in quasi-elastic
electron scattering on nuclei is actually the light cone plus-momentum
distribution function for nucleons in the nucleus. It is useful to use
a relativistic form of the variable $y$ \cite{sdm} in which
\begin{equation}
y=-q^3+ \nu+E_s+M,
\end{equation}
as both $q^3$ and $\nu$ are large in magnitude, and $E_s$ is the single
nucleon separation energy. But $x ={(q^3)^2-\nu^2\over 2M\nu}\approx
1+{E_s-y\over M}$, so that
\begin{equation}
Mx=M+E_s-y\equiv M_Ay_A.
\end{equation}
Here $y_A$ is a new y-scaling variable. This means that according to
Eq.~(\ref{one})
\begin{equation}
p_B^+=M_A y_A/M\approx Ay_A,
\end{equation}
so that a measurement of $\sigma$ determines the probability that the
struck nucleon has a plus-momentum of $Ay_A$. This probability also
enters in convolution model calculations of nuclear deep inelastic
scattering.

The use of light-front dynamics to compute nuclear wave functions
should allow us to compute $F(y)$ from first principles. Furthermore,
we claim that using light-front dynamics incorporates the
experimentally relevant kinematics from the beginning, and therefore is
the most efficient way to compute the cross sections for nuclear deep
inelastic scattering and nuclear quasi-elastic scattering. 

It is worthwhile to review some of the features of the EMC
effect \cite{emc,emcrevs}. The key experimental result is the
suppression of the structure function for $x\sim 0.5$. This means that
the valence quarks of bound nucleons carry less plus-momentum than
those of free nucleons. One way to understand this result is to
postulate that mesons carry a larger fraction of the plus-momentum in
the nucleus than in free space. While such a model explains the shift
in the valence distribution, one obtains at the same time a meson (i.e.
anti-quark) distribution in the nucleus, which is strongly enhanced
compared to free nucleons and which should be observable in Drell-Yan
experiments \cite{dyth}. However, no such enhancement has been observed
experimentally \cite{dyexp}, and the implications are analyzed in
Ref.~\cite{missing}.

The use of light-front dynamics allows us to compute the necessary
nuclear meson distribution functions using variables which are
experimentally relevant. The need for a computation of such functions
in a manner consistent with generally known properties of nuclei led
one of us to attempt to construct a light front treatment of nuclear
physics \cite{jerry}. These calculations, using a Lagrangian in which
Dirac nucleons are coupled to massive scalar and vector
mesons \cite{bsjdw}, treated the example of infinite nuclear matter
within the mean field approximation. In this case, the meson fields are
constants in both space and time and the momentum distribution has
support only at $k^+=0$. Such a distribution would not be accessible
experimentally, so that the suppression of the plus-momentum of valence
quarks would not imply the existence of a corresponding testable
enhancement of anti-quarks. However, it is necessary to ask if the
result is only a artifact of the infinite nuclear size and of the mean
field approximation. The present paper is an attempt to handle
finite-sized nuclei using light-front dynamics.

\subsection{Recovery of rotational invariance}

It is worthwhile to discuss, in a general way, how it is that we are
able to find spectra which have the correct number of degenerate
states. Let us imagine that we try to determine eigenstates of a LF
Hamiltonian by means of a variational calculation. Simply minimizing
the LF energy obviously leads to nonsensical results since the LF
energy scales like the inverse of the LF momentum. Even if one has only
a poor ansatz for the intrinsic wave function, one can easily reach
zero energy by letting the overall momentum scale to infinity! However,
this problem is avoided by performing a constrained variation, in which
the total LF momentum is fixed by including a Lagrange multiplier term
proportional to the total momentum in the LF Hamiltonian. Note that
this is not a problem if one is able to use a Fock space basis in which
the total plus and $\perp$ momentum of each component are fixed. In
calculations involving many particles, the Fock state approach cannot
be used in practical calculations --- instead one uses a mean field in
which each particle moves in an ``external'' potential. In this case
the total momentum is not fixed, and a Lagrange multiplier term needs
to be included in order to avoid solutions with infinite LF momentum.

In order to fix this potential problem with ``runaway solutions''
($P^+\to \infty$) to variational calculations for LF
Hamiltonians, any term proportional to $P^+$ would suffice. However, by
setting the coefficient for the term proportional to $P^+$ equal to
one, {\it i.e.} minimizing $P^-+P^+$, one automatically guarantees that
$P^+=P^-$ (or $P^3=0$). The reason is that, using covariance, $P^-$ has
eigenvalues of the form $P^-_n =\frac{M^2_n+P_\perp^2}{P^+}$,
{\it i.e.} it scales like $1/P^+$. Therefore, when one minimizes $P^- +
P^+$ with respect to $P^+$, the minimum occurs for $P^+ =
\sqrt{{M^2_n+P_\perp^2}}$, which yields $P^- ={M^2_n+{
P_\perp}^2\over{P^+}} = \sqrt{{M^2_n+P_\perp^2}}$ as
well. This ``equipartition'' between $P^+$ and $P^-$ thus arises since
the two operators scale in exactly opposite ways under longitudinal 
boosts. Note that this is quite analogous to the nonrelativistic
harmonic oscillator where, under scale transformations, potential and
kinetic energy scale in opposite ways, resulting in the equipartition
between potential and kinetic energy.

The net result is that we minimize the sum of $P^++P^-$. The need to
include the plus-momentum can also be seen in a simple example.
Consider a nucleus of $A$ nucleons of momentum $P_A^+=M_A$,
${\bbox{P}_A}_\perp=0$, which consists of a nucleon of momentum
$(p^+,\bbox{p}_\perp)$, and a residual $(A-1)$ nucleon system which
must have momentum $(P^+_A-p^+,-\bbox{p}_\perp)$. The kinetic energy
$K$ is given by the expression
\begin{equation}
K={p_\perp^2+M^2\over p^+}+{p_\perp^2+M_{A-1}^2\over P^+_A- p^+}.
\end{equation}
In the second expression, one is tempted to neglect the term $p^+$ in
comparison with $ P^+_A\approx M_A$. This would be a mistake. Instead
make the expansion
\begin{eqnarray}
K&\approx&{p_\perp^2+M^2\over p^+}+{M_{A-1}^2\over P^+_A}\left(1+ {p^+\over
 P_A^+}\right)\nonumber\\ 
&\approx&{p_\perp^2+M^2\over p^+}+p^+ +M_{A-1}, 
\end{eqnarray}
because for large $A$, $M^2_{A-1}/P_A^2\approx 1$. For free particles,
of ordinary three momentum $\bbox{p}$ one has $E^2(p)=\bbox{p}^2+m^2$
and $p^+=E(p)+p^3$, so that
\begin{equation}
K\approx {\left(E^2(p)-(p^3)^2\right)\over E(p)+p^3}
+E(p)+p^3+M_{A-1}=2E(p)+M_{A-1}.
\end{equation}
We see that $K$ depends only on the magnitude of a three-momentum and
rotational invariance is restored. The physical mechanism of this
restoration is the inclusion of the recoil kinetic energy of the
residual nucleus.

\subsection{Outline}

The organization of the paper is as follows. The light-front
quantization for our chosen Lagrangian is is presented in Sec.~II. This
quantization is applied, along with a constrained minimization of the
expectation value of $P^-$, to derive a light-front version of mean
field theory in Sec.~III. We obtain a new light version of the equation
that defines the single nucleon modes. The solutions of this equation
are approximately a non-trivial phase factor times the solutions of the
usual equal-time ET Dirac equation. The consequences of this phase
factor are discussed. 

The meson fields are treated as expectation values of operators. The
equations for these expectation values are closely related to the meson
field equations appearing in the usual treatment of the Walecka model.
However, the mesonic Fock space is accessible in our formalism. Our
nucleon mode equation is simplified by the use of a two-component
spinor formalism \cite{harizhang}, and by an angular momentum reduction
in Sec.~IV. The numerical aspects are discussed in App.~\ref{appa}. The
binding energies, nucleon and meson distributions for $^{16}$O and
$^{40}$Ca are presented in Sec.~V. A concluding discussion appears in
Sec.~VI. Numerical details of how we evaluate the momentum
distributions are given in App.~\ref{appb}. A brief discussion of some
of the results can be found in Ref.~\cite{us}. A related set of
solutions of some toy model problems and a heuristic derivation of our
nucleon mode equation will appear in a separate  paper \cite{us2}.

\section{Light-Front Quantization}

We start with a model in which the nuclear constituents are nucleons
$\psi$ (or $\psi')$, scalar mesons $\phi$ and vector mesons $V^\mu$.
The Lagrangian ${\cal L}$ is given by 
\begin{equation}
{\cal L} ={1\over 2} (\partial^\mu \phi \partial_\mu \phi-m_s^2\phi^2) 
-{1\over 4} V^{\mu\nu}V_{\mu\nu} +{1\over 2}m_v^2V^\mu V_\mu
+\bar{\psi}^\prime\left(\gamma^\mu
({i\over 2}\stackrel{\leftrightarrow}{\partial}_\mu
-g_v\;V_\mu) -
M -g_s\phi\right)\psi', \label{lag}
\end{equation}
where the bare masses of the nucleon, scalar and vector mesons are
given by $M, m_s,$ $m_v$, and $V^{\mu\nu}= \partial ^\mu
V^\nu-\partial^\nu V^\mu$. We ignore pions here.

The field equations are given by:
\begin{eqnarray}
\gamma\cdot(i\partial-g_v V)\psi'&=&(M\;+g_s\phi)\psi',
\label{dirac1}\\
\partial_\mu V^{\mu\nu}+m_v^2 V^\nu&=&g_v\bar \psi'\gamma^\nu\psi'
\label{vmeson}\\
\partial_\mu\partial^\mu \phi+m_s^2\phi&=&-g_s\bar\psi'\psi'. \label{smeson}
\end{eqnarray}

The next step is obtain the light-front Hamiltonian
($P^-$) \cite{notation} as a sum of a free, non-interacting and a set of
terms containing all of the interactions. This is accomplished by
separating the independent and dependent degrees of freedom in the
usual way \cite{lcrevs,mb:adv} and then using the energy momentum
tensor. Consider the nucleons. Although described by four-component
spinors, these fields have only two independent degrees of freedom. 
The light-front formalism allows a convenient separation of dependent
and independent variables via the projection operators
$\Lambda_\pm\equiv \case{1}{2}\gamma^0\gamma^\pm$ \cite{des71,yan12},
with $\psi'_\pm\equiv\Lambda_\pm \psi'$. The independent fermion degree
of freedom is chosen to be $\psi'_+$, and one finds
\begin{eqnarray}
(i\partial^--g_vV^-)\psi'_+&=&(\bbox{\alpha}_\perp\cdot
(\bbox{p}_\perp-g_v\bbox{V}_\perp)+\beta (M +g_s\phi))\psi'_-\nonumber\\
(i\partial^+-g_vV^+)\psi'_-&=&(\bbox{\alpha}_\perp\cdot
(\bbox{p}_\perp-g_v\bbox{V}_\perp)+\beta (M +g_s\phi))\psi'_+.\label{nfg}
\end{eqnarray}
The relation between $\psi'_-$ and $\psi'_+$ is very complicated
unless one may set the plus component of the vector field to
zero \cite{lcrevs}. This is a matter of a choice of gauge for QED and
QCD, but the non-zero mass of the vector meson prevents such a choice 
here. Instead, one simplifies the equation for $\psi'_-$
by \cite{des71,yan34} transforming the fermion field according to 
\begin{equation}
\psi'=e^{-ig_v\Lambda(x)}\psi ,\qquad
\partial^+ \Lambda=V^+ \label{trans}.
\end{equation}
This transformation leads to the replacement of Eq.~(\ref{nfg}) by 
\begin{eqnarray}
(i\partial^--g_v \bar V^-)\psi_+&=&(\bbox{\alpha}_\perp\cdot 
(\bbox{p}_\perp-g_v\bbox{\bar V}_\perp)+\beta(M+g_s\phi))\psi_-\nonumber\\
i\partial^+\psi_-&=&(\bbox{\alpha}_\perp\cdot 
(\bbox{p}_\perp-g_v\bbox{\bar V}_\perp)+\beta(M+g_s\phi))\psi_+,\label{yan}
\end{eqnarray}
where
\begin{equation}
\partial^+\bar V^\mu=\partial^+V^\mu-\partial^\mu V^+.
\label{vbar}
\end{equation}
Note that while it is $\bar V^\mu$ that enters in the nucleon field
equations, it is $V^\mu$ that enters in the meson field equations. 

The scalar field can be expressed in terms of creation and destruction 
operators:
\begin{equation}
\phi(x)=
\int{ d^2k_\perp dk^+ \theta(k^+)\over (2\pi)^{3/2}\,\sqrt{2k^+}}\left[
a(\bbox{k})e^{-ik\cdot x}
+a^\dagger(\bbox{k})e^{ik\cdot x}\right],\label{phiex}
\end {equation} 
where
\begin{equation}
k\cdot x={1\over2}(k^+x^-)-\bbox{k}_\perp\cdot \bbox{x}_\perp\equiv
\bbox{k}\cdot\bbox{x},
\label{kx}
\end{equation}
and the fields and their derivatives with respect to $x^+$ are
evaluated at $x^+=0$. This notation is used through out this work. The
consequence is that the energy momentum tensor $T^{\mu\nu}$ does not
depend on $x^+$. In the above expansion (and in the expansions for any
of our fields) the particles are on the mass-shell. Here 
$k^-={k_\perp^2+m_s^2\over k^+}$. The theta function restricts $k^+$
to positive values. The commutation relations are
\begin{equation}
[a(\bbox{k}),a^\dagger(\bbox{k}')]=
\delta(\bbox{k}_\perp-\bbox{k}'_\perp)
\delta(k^+-k'^+),\label{comm}
\end {equation}
with $[a(\bbox{k}),a(\bbox{k}')]=0$.
It is useful to define 
\begin{equation}
\delta^{(2,+)}(\bbox{k}-\bbox{k}')\equiv 
\delta(\bbox{k}_\perp-\bbox{k}'_\perp)\delta(k^+-k'^+).\label{def3d}
\end {equation} 

The expression for the vector meson field operator is
\begin{equation}
V^\mu(x)=
\int{ d^2k_\perp dk^+ \theta(k^+)\over (2\pi)^{3/2}\,\sqrt{2k^+}}
\sum_{\omega=1,3}\epsilon^\mu(\bbox{k},\omega)\left[
a(\bbox{k},\omega)e^{-i\bbox{k}\cdot\bbox{x}}
+a^\dagger(\bbox{k},\omega)e^{i\bbox{k}\cdot\bbox{x}}\right],\label{vfield}
\end {equation} 
where the polarization vectors are the usual ones:
\begin{eqnarray}
k^\mu\epsilon_\mu(\bbox{k},\omega)=0,\qquad \epsilon^\mu(\bbox{k},\omega)
\epsilon_\mu(\bbox{k},\omega')=-\delta_{\omega\omega'},\nonumber\\
\sum_{\omega=1,3}\epsilon^\mu(\bbox{k},\omega)
\epsilon^\nu(\bbox{k},\omega)=-(g^{\mu\nu}-{k^\mu k^\nu\over m_v^2}).
\label{p1}
\end{eqnarray}
Once again the four momenta are on-shell, with 
$k^-={k_\perp^2+m_v^2\over k^+}.$ The commutation relations are
\begin{equation}
[a(\bbox{k},\omega),a^\dagger(\bbox{k}',\omega')]=\delta_{\omega\omega'}
\delta^{(2,+)}(\bbox{k}-\bbox{k}'), 
\end{equation}
with $[a(\bbox{k},\omega),a(\bbox{k}',\omega')]=0$,
and lead to commutation relations amongst the
field operators that are the 
same as in Ref.~\cite{yan34}.

We also need the eigenmode expansion for $\bar V^\mu$. This is given by
\begin{equation}
\bar V^\mu(x)=
\int{ d^2k_\perp dk^+ \theta(k^+)\over (2\pi)^{3/2}\,\sqrt{2k^+}}
\sum_{\omega=1,3}\bar\epsilon^\mu(\bbox{k},\omega)\left[
a(\bbox{k},\omega)e^{-ik\cdot x}
+a^\dagger(\bbox{k},\omega)e^{ik\cdot x}\right],\label{nvfield}
\end {equation} 
where, using Eqs.(\ref{vbar}) and (\ref{vfield}),
the polarization vectors $\bar\epsilon^\mu(\bbox{k},\omega)$ are
\begin{equation}
\bar\epsilon^\mu(\bbox{k},\omega)=
\epsilon^\mu(\bbox{k},\omega)-{k^\mu\over k^+}\epsilon^+(\bbox{k},\omega).
\label{p2}
\end{equation}
Note that
\begin{equation}
\sum_{\omega=1,3}\bar\epsilon^\mu(\bbox{k},\omega)
\bar\epsilon^\nu(\bbox{k},\omega)=-(g^{\mu\nu}-g^{+\mu}{k^\nu\over k^+}
-g^{+\nu}{k^\mu\over k^+}).
\label{ebar}
\end{equation}

Then we may construct the total four-momentum operator from 
\begin{equation}
P^\mu={1\over2}\int dx^-d^2x_\perp\,T^{+\mu}(x^+=0,x^-,\bbox{x}_\perp),
\end{equation}
with (as usual)
\begin{equation}
T^{\mu\nu}=-g^{\mu\nu}{\cal L} +\sum_r{\partial{\cal L}\over\partial
(\partial_\mu\phi_r)}\partial^\nu\phi_r\, ,\label{tmunu}
\end {equation}
in which the degrees of freedom are labelled by $\phi_r$. We need
$T^{++}$ and $T^{+-}$, which are 
\begin{equation}
T^{++}=\partial^+\phi\partial^+\phi+V^{ik}V^{ik}
+m_v^2V^+ V^+
+2\psi^\dagger_+ i\partial^+ \psi_+,
\label{tpp}
\end{equation}
and 
\begin{eqnarray}
T^{+-}&=&\bbox{\nabla}_\perp\phi\cdot\bbox{\nabla}_\perp\phi +m_\phi^2\phi^2
+{1\over 4}(V^{+-})^2+{1\over 2}V^{kl}V^{kl} +m^2_vV^kV^k\nonumber\\
&&+\bar\psi
\left(\bbox{\gamma}_\perp\cdot(\bbox{p_\perp}-g_v\bbox{\bar{V}^-})
+M+g_s\phi\right)\psi.
\label{tpm2}
\end{eqnarray}

This form is still not useful for calculations because the constrained
field $\psi_-$ contains interactions. We follow Refs.~ \cite{des71,mpsw}
in expressing $\psi_-$ as a sum of terms, one $\xi_-$ whose relation
with $\psi_+$ is free of interactions, the other $\eta_-$ containing
the interactions. That is, rewrite the second of Eq.~(\ref{yan}) as
 \cite{harizhang}
\begin{eqnarray}
\xi_-&=&{1\over i\partial^+}(\bbox{\alpha}_\perp\cdot 
\bbox{p}_\perp+\beta M)\psi_+\nonumber\\
\eta_-&=&{1\over i\partial^+}(-\bbox{\alpha}_\perp\cdot 
g_v\bbox{\bar V}_\perp+\beta g_s\phi)\psi_+. \label{yan1}
\end{eqnarray}
Furthermore, define $\xi_+(x)\equiv\psi_+(x)$, so that 
\begin{equation}
\psi(x)=\xi(x)+ \eta_-(x), \label{fcon}
\end{equation}
where $\xi(x)\equiv \xi_-(x)+\xi_+(x)$. This separates the dependent
and independent parts of $\psi$.

One needs to make a similar treatment for the vector meson fields. The
operator $V^{+-}$, is determined by
\begin{equation}
V^{-+}={2\over\partial^+}\left[g_v\;J^+-m^2_vV^+-\partial_iV^{i+}\right].
\label{vpm}
\end{equation} 
Part of this operator is determined by a constraint equation, because
the independent variables are $V^+$ and $V^{i+}$. To see this examine
Eq~(\ref{vpm}), and make a definition
\begin{equation}
V^{+-}=v^{+-}+\omega^{+-}, \label{vcon}
\end{equation}
where 
\begin{equation}
\omega^{+-}=-{2\over \partial^+}J^+.
\end{equation}
The sum of the last term of Eq\.~(\ref{tpm2}) and the terms involving
$\omega^{+-}$ is the interaction density. Then one may use
Eqs.~(\ref{tpm2}), (\ref{fcon}), and (\ref{vcon}) to rewrite the $P^-$
as a sum of different terms, with 
\begin{equation}
P_{0N}^-={1\over 2}\int d^2x_\perp dx^-\,
\bar\xi\left(\bbox{\gamma}_\perp\cdot\bbox{p}_\perp +M\right)\xi,
\label{freef}
\end{equation}
and the interactions
\begin{equation}
P^-_I=v_1+v_2+v_3, \label{defv}
\end{equation}
with 
\begin{equation}
v_1=\int d^2x_\perp dx^-\,\bar\xi\left(g_v\gamma\cdot\bar V+g_s\phi
\right)\xi,\label{v1}
\end{equation}
\begin{equation}
v_2=\int d^2x_\perp dx^-\,\bar\xi\left(-g_v\gamma\cdot\bar V
+g_s\phi\right)\;{\gamma^+\over 2i\partial^+}\;\left(-g_v\gamma\cdot\bar V
+g_s\phi
\right)\xi, \label{v2}
\end{equation}
and 
\begin{eqnarray}
v_3&=&{g_v^2\over8}\int d^2x_\perp dx^-\int dy^-_1\,
\epsilon(x^--y^-_1)\,
\xi^\dagger_+(y^-_1,\bbox{x}_\perp)\gamma^+\xi_+(y^-_1,\bbox{x}_\perp)
\nonumber\\
&&\times \int dy^-_2\,\epsilon(x^--y^-_2)
\xi^\dagger_+
(y^-_2,\bbox{x}_\perp)\gamma^+\xi_+(y^-_2,\bbox{x}_\perp),\label{v3}
\end{eqnarray}
where $\epsilon(x)\equiv \theta(x)-\theta(-x)$. The term $v_1$ accounts
the emission or absorption of a single vector or scalar meson. The term
$v_2$ includes contact terms in which there is propagation of an
instantaneous fermion. The term $v_3$ accounts for the propagation of
an instantaneous vector meson.

Our variational procedure will involve the independent fields $\psi_+$,
so we need to express the interactions $P_{0N}^-, v_{1,2}$ in terms of
$\xi_+$. A bit of Dirac algebra shows that
\begin{eqnarray}
P^-_N&\equiv& P^-_{0N}+v_1+v_2\nonumber\\
&=&
\int d^2x_\perp {dx^-\over 2}\,\xi^\dagger_+\Bigl[2g_v\bar{V}^-
\nonumber\\
&+& \left(\bbox{\alpha}_\perp
\cdot(\bbox{p}_\perp-g_v\bbox{\bar V}_\perp)+\beta(M+g_s\phi)\right)
{1\over i\partial^+}
\left(\bbox{\alpha}_\perp
\cdot(\bbox{p}_\perp-g_v\bbox{\bar V}_\perp)
+\beta(M+g_s\phi)\right)\Bigr]\xi_+.
\label{good}
\end{eqnarray}

It is worthwhile to define the contributions to $P^\pm$ arising from
the mesonic terms as $P_s^\pm$ and $P_v^\pm$. Then one may use
Eqs.~(\ref{tpm2}) and (\ref{tpp}) along with the field expansions to
obtain
\begin{eqnarray}
P_s^-&=&{1\over2}\int d^2x_\perp dx^-\,
\left(\bbox{\nabla}_\perp\phi\cdot\bbox{\nabla}_\perp\phi +m_s^2\phi^2
\right)\nonumber \\
&=&\int d^2k_\perp dk^+\,\theta(k^+)a^\dagger
(\bbox{k})a(\bbox{k}){k_\perp^2+m_s^2\over k^+},
\label{pmph}
\end{eqnarray}
\begin{equation}
P_s^+=\int d^2k_\perp dk^+\,\theta(k^+)a^\dagger
(\bbox{k})a(\bbox{k})k^+,
\label{ppph}
\end{equation}
\begin{equation}
P_v^-=\sum_{\omega=1,3}\int{ d^2k_\perp dk^+ \,
\theta(k^+){k_\perp^2+m_v^2\over k^+}} 
a^\dagger(\bbox{k},\omega) a(\bbox{k},\omega)\;+v_3,\label{pvm}
\end{equation}
and
\begin{equation}
P_v^+=\sum_{\omega=1,3}\int d^2k_\perp dk^+ \,\label{ppm}
\theta(k^+)k^+
a^\dagger(\bbox{k},\omega) a(\bbox{k},\omega).
\end {equation}
The term $v_3$ is the vector-meson instantaneous term, and we include
it together with the purely mesonic contribution to $P_v^-$ because it
is cancelled by part of that contribution.

Thus, our result for the total minus-momentum operator is
\begin{equation}
P^-=P_N^- +P_s^- + P_v^-,
\label{bigm}
\end{equation}
and for the plus-momentum
\begin{equation}
P^+=P_N^+ +P_s^+ + P_v^+
,\label{bigp}
\end{equation}
where from Eq.~(\ref{tpp})
\begin{equation}
P_N^+\equiv \int d^2x_\perp {dx^-\over 2}\, 2\xi_+^\dagger i\partial^+\xi_+.
\end{equation}

\section{Mean Field Theory}

The light-front Schroedinger equation for the complete nuclear
ground-state wave function $\mid\Psi\rangle$ is
\begin{equation}
 P^-\mid\Psi\rangle = M_A\mid\Psi\rangle.
\end{equation}
We choose to work in the nuclear rest frame so that we also need
\begin{equation}
 P^+\mid\Psi\rangle = M_A\mid\Psi\rangle.
\end{equation}
We want to use a variational principle. One might think that one may
simply minimize the expectation value of $P^-$, but this makes no sense
because $P^+P^-=M_A^2$ when acting on the wave function. One would get
a zero of $P^-$ for an infinite value of $P^+$. As explained in the
Introduction, one must minimize the expectation value of $P^-$ subject
to the condition that the expectation value of $P^+$ is equal to the
expectation value of $P^-$. This is the same as minimizing the average
of $P^-$ and $P^+$, which is the rest-frame energy of the entire
system. To this end we define a light-front Hamiltonian
\begin{equation}
 H_{LF}\equiv {1\over 2}\left(P^++P^-\right).
\end{equation} 
We stress that $H_{LF}$ is not usual the Hamiltonian, because the
light-front quantization is used to define all of the operators that
enter. 

The wave function $\mid\Psi\rangle$ consists of a Slater determinant of
nucleon fields $\mid\Phi\rangle$ times a mesonic portion
\begin{equation}
\mid\Psi\rangle= \mid\Phi\rangle\otimes\mid \rm{mesons}\rangle,
\end{equation}
and the mean field approximation is characterized by the replacements
\begin{eqnarray}
\phi&\to& \langle\Psi\mid \phi\mid\Psi\rangle\nonumber\\
V^\mu&\to& \langle\Psi\mid V^\mu\mid\Psi\rangle. \label{replace1}
\end{eqnarray}
We shall derive the meson field equations, and then determine the
nucleon modes using a variational principle.

\subsection{Meson field equations}

We shall go through the derivation of the equation for the expectation
value of $\phi(x)$ in a detailed fashion. Consider the quantity $H_{LF}
a(\bbox{k})\mid \Psi\rangle$, and use commutators to obtain
\begin{equation}
H_{LF}a(\bbox{k})\mid \Psi\rangle=[ H_{LF},a(\bbox{k}) ]\mid \Psi\rangle
+M_A a(\bbox{k})\mid \Psi\rangle.\label{ha}
\end{equation}
The operators $P_s^\pm$ of Eqs.~(\ref{pmph}) and (\ref{ppph})
and the standard commutation relations allow one to obtain
\begin{equation}
[ H_{LF},a(\bbox{k}) ]=-{k_\perp^2+{k^+}^2+m_s^2\over 2k^+}
a(\bbox{k}) +
{J(\bbox{k})\over (2\pi)^{3/2}\sqrt{2k^+}},
\end{equation}
where ${J(\bbox{k})\over (2\pi)^{3/2}\sqrt{2k^+}}$ is the commutator of
the interaction, Eq.~(\ref{defv}), between the scalar meson and the
nucleon:
\begin{equation}
{J(\bbox{k})\over (2\pi)^{3/2}\sqrt{2k^+}}={1\over 2}
[P^-_I,a((\bbox{k}) ].
\end{equation} 

We use Eqs.~(\ref{v1})-(\ref{v3}), and take the commutator of the
interactions $v_i$ with $a(\bbox{k})$. Then re-express the results in
terms of $\xi$ to obtain
\begin{equation}
J(\bbox{k}) =-{1\over 2}g_s\int
d^2x_\perp dx^-\, e^{i\bbox{k}\cdot \bbox{x}}
\bar\xi(\bbox{x})\xi(\bbox{x}) . \label{jk}
\end{equation}
Take the overlap of Eq.~(\ref{ha}) with $\langle \Psi\mid$ to find
\begin{equation}
\langle \Psi\mid a(\bbox{k})\mid \Psi\rangle
{k_\perp^2+{k^+}^2+m_s^2\over 2k^+}
={ \langle \Psi\mid J(\bbox{k})\mid \Psi\rangle 
\over (2\pi)^{3/2}\sqrt{2k^+}}\;. \label{pap}
\end{equation}
Multiply the above Eq.~(\ref{pap}) by a factor ${\sqrt{2k^+}\over
(2\pi)^{3/2}}e^{-i\bbox{k}\cdot \bbox{x}}$. Then add the result of that
operation to its complex conjugate. The integral of the resulting
equation over all $\bbox{k}_\perp$ and positive values of $k^+$ and
using the field expansion (\ref{phiex}) leads to the result
\begin{eqnarray}
\lefteqn{
\left(-\nabla_\perp^2-
\left(2{\partial \over \partial x^-}\right)^2+m_s^2 
\right)
\langle \Psi\mid \phi(x)\mid \Psi\rangle =}
\qquad\qquad&&\nonumber \\
&&\langle \Psi\mid 
\int {d^2k_\perp dk^+\theta(k^+)\over (2\pi)^3}\,\left(J(\bbox{k})
e^{-i\bbox{k}\cdot \bbox{x}} +J^\dagger(\bbox{k})
e^{+i\bbox{k}\cdot \bbox{x}}\right)\mid \Psi\rangle .
\label{ph1}
\end{eqnarray}

The evaluation of the right-hand-side of Eq.~(\ref{ph1}) proceeds by
using Eq.~(\ref{jk}) and its complex conjugate. The combination of
those two terms allows one to remove the factor $\theta(k^+)$ and
obtain a delta function from the momentum integral. That ${1\over2}k^+$
appears in the exponential leads to the removal of the factor
${1\over2}$ of Eq.~(\ref{jk}). One can also change variables using
\begin{equation}
z\equiv{-x^-\over 2},\qquad \bbox{x}\equiv (z,\bbox{x}_\perp).
\label{relate}
\end{equation}
The minus sign enters to remove the minus sign between the two terms of
the factor $k\cdot x$ in Eq.~(\ref{kx}). Then one may re-define the
operator $-\nabla_\perp^2- \left(2{\partial \over \partial
x^-}\right)^2$ appearing in Eq.~(\ref{ph1}) as $-\nabla^2$. Note that
we previously \cite{bm98} obtained the above relation \ref{relate}
simply by examining the space-time diagram for a static source
(independent of $x^0$). The net result is that
\begin{equation}
\left(-\nabla^2+m_s^2 \right)
\langle \Psi\mid \phi(\bbox{x})
\mid \Psi\rangle =-g_s \langle \Psi\mid
\bar{\psi}
(\bbox{x})\psi(\bbox{x})\mid \Psi\rangle ,\label{phieq}
\end{equation}
which has the same form as the equation in the usual equal-time
formulation. Note that the right hand side of Eq.~(\ref{phieq}) should
be a function of $|\bbox{x}|$ for the spherical nuclei of our
present concern. Our formalism for the nucleon fields uses
$\bbox{x}_\perp$ and $x^-$ as independent variables, so that obtaining
numerically scalar and vector nucleon densities that depend only
$x_\perp^2+(x^-/2)^2$ will provide a central, vital test of our
procedures and mean field theory. Assuming for the moment that this
occurs, the scalar field $\langle\Psi\mid
\phi(\bbox{x})\mid\Psi\rangle$ will depend only $|\bbox{x}|$
according to (\ref{phieq}).

We stress that the use of Eq.~(\ref{relate}) is merely a convenient way
to simplify the calculation --- using it allows us to treat the $\perp$
and minus spatial variables on the same footing, and to maintain
explicit rotational invariance. We will obtain the mesonic
plus-momentum distributions from the ground state expectation value of
different operators.

The procedure of Eqs.~(\ref{ha}) to (\ref{phieq}) can also be applied
to the vector fields. The appearance of the barred vector potential
makes it necessary to display certain steps. The starting point is to
consider the expression $H_{LF}a(\bbox{k})\mid \Psi\rangle$ and the
interaction
\begin{equation}
{J(\bbox{k},\omega)\over (2\pi)^{3/2}\sqrt{2k^+}}={1\over 2}
[P^-_I,a(\bbox{k},\omega) ]. 
\end{equation} 
Using equations (\ref{v1})-(\ref{v3}), taking the commutator of the
interactions $v_i$ with $a(\bbox{k})$, and re-expressing the results
in terms of $\xi$, leads to
\begin{equation}
J(\bbox{k},\omega) =-{1\over 2}g_v\int
d^2x_\perp dx^-\, e^{ik\cdot x}
\bar\xi(\bbox{x})\gamma\cdot\bar{\epsilon}(\bbox{k},\omega) \xi(\bbox{x})
. \label{jwk}
\end{equation}
This, along with the other terms in the expression for
$H_{LF}a(\bbox{k},\omega)\mid \Psi\rangle$, allows us to obtain
\begin{equation}
\langle \Psi\mid a(\bbox{k},\omega)\mid \Psi\rangle
{k_\perp^2+{k^+}^2+m_v^2\over 2k^+}
={ \langle \Psi\mid-{1\over 2}g_v\int
d^2x_\perp dx^-\, e^{ik\cdot x}
\bar\xi(\bbox{x})\gamma\cdot\bar{\epsilon}(\bbox{k},\omega) \xi(\bbox{x})
\mid \Psi\rangle 
\over (2\pi)^{3/2}\sqrt{2k^+}}. \label{pvp}
\end{equation}
The field equation for the vector mesons $\bar V^\mu$ is obtained by
multiplying the above by $\bar{\epsilon}^\mu(\bbox{k},\omega)$, summing
over $\omega$ and performing standard manipulations. We need to know
the quantity
\begin{equation}
X^\mu(\bbox{k})\equiv \sum_{\alpha,\omega}\gamma_\alpha\bar{\epsilon}^\alpha
(\bbox{k},\omega)\bar{\epsilon}^\mu(\bbox{k},\omega)
\end{equation} The use of Eqs.~(\ref{ebar}) and (\ref{p2}) leads to
\begin{equation}
X^\mu(\bbox{k})=
-\gamma^\mu+2\delta(\mu,-){\gamma\cdot k\over k^+}
+{k^\mu\over k^+}\gamma^+.
\label{xmu}
\end{equation}
One makes familiar manipulations to obtain the result
\begin{equation}
\left(-\nabla^2+m_v^2\right)
\langle \Psi\mid \bar{V}^\mu (\bbox{x})
\mid \Psi\rangle =g_v \langle \Psi\mid
\bar\xi(\bbox{x})\gamma^\mu\xi(\bbox{x})\mid \Psi\rangle
+\Delta^\mu,\label{vbeq}
\end{equation}
with
\begin{equation} 
\Delta^\mu(\bbox{x})\equiv- \langle \Psi\mid
\int{ d^3x'\over(2\pi)^3}\, e^{i\bbox{k}\cdot
(\bbox{x}-\bbox{x}')} \bar\xi(\bbox{x}')
\left({k^\mu\over k^+}\gamma^+
+ 2\delta(\mu,-){\gamma\cdot k\over k^+}\right)\xi(\bbox{x}')\mid\Psi.\rangle
\label{delta}
\end{equation}

Note that (as in the derivation given above for $\phi$) the variable
$k^+$ (confined to positive values) is replaced by the inclusion of the
complex conjugate term $(a^\dagger(\bbox{k},\omega) $ by a variable
$k^3$ which ranges from $-\infty$ to $\infty$. We proceed by first
assuming that
\begin{equation}
\left(-\nabla^2+m_v^2\right)
\langle \Psi\mid {V}^\mu (\bbox{x})
\mid \Psi\rangle =g_v \langle \Psi\mid
\bar\xi(\bbox{x})\gamma^\mu\xi(\bbox{x})\mid \Psi\rangle\equiv J^\mu,
\label{veq}
\end{equation}
which is to be verified by showing that Eqs.~(\ref{vbeq}) and
(\ref{veq}) are consistent with the defining relation (\ref{vbar})
Taking the difference between Eqs.~(\ref{vbeq}) and (\ref{veq}) and
using the defining relation leads \cite{cc} to a consistency
requirement
\begin{equation} 
\partial^\mu J^+=\partial^+\Delta^\mu. \label{equal}
\end{equation}
For $\mu\ne-$ the above relation is verified by integration by parts in
the expression (\ref{delta}) for $\Delta^\mu$. If $\mu=-$, one may use
the definition (\ref{vbar}) and that $\langle\Psi\mid
V^+\mid\Psi\rangle$ does not depend on $x^+$ to see that
\begin{equation}
\partial^-\langle\Psi\mid\bar {V}^-\mid\Psi\rangle
=\partial^-\langle\Psi\mid V^-\mid\Psi\rangle\label{vm}.
\end{equation}
Thus the validity of Eq.~(\ref{veq}) is established. 

\subsection{Nucleon single-particle wave functions}

The mesonic field equations are given in the previous subsection. The
equation for the nucleon modes are to be found using the procedure of
minimizing $P^-+P^+$ with respect to the nucleon wave function, subject
to the condition that the normalization of the independent fields
remains fixed. The nucleon field operators enter only in the term
$P_N^-+P_N^+$, so that it is useful to define
\begin{equation}
H_{LF}\equiv {1\over 2}\left(P_N^-+P_N^+\right).
\end{equation}
The specific operator is obtained by using Eq.~(\ref{good}) to find
\begin{equation}
H_{LF}=\int {dx^-\over 2}
d^2x_\perp\, \xi^\dagger_+ {\cal H}_{LF}\xi_+,
\end{equation}
where
\begin{eqnarray}
2{\cal H}_{LF}&\equiv& 
i\partial^++ 2g_v\bar{V}^- \nonumber\\
& +& \left(\bbox{\alpha}_\perp
\cdot(\bbox{p}_\perp-g_v\bbox{\bar V}_\perp)+\beta(M+g_s\phi)\right)
{1\over i\partial^+}
\left(\bbox{\alpha}_\perp
\cdot(\bbox{p}_\perp-g_v\bbox{\bar V}_\perp)
+\beta(M+g_s\phi)\right). \label{hlf}
\end{eqnarray}
The potentials appearing in Eq.~(\ref{hlf}) are independent of $x^+$.
This implies some simplifications: $\partial^+ \bar {V}^-=\partial^+
V^--\partial^-V^+=\partial^+ V^-$, so that $\bar V^-=V^-$, and (for
$i=1,2$) $\partial^+ \bar {V}^i=\partial^+ V^i-\partial^i
V^+=-\partial^i V^+$. Using the relation (\ref{trans}) we find that
$\bar {V}^i=-\partial^i\Lambda$.

The Slater determinant $\mid\Phi\rangle$ is defined by allowing $A$
nucleon states, denoted by the index $\alpha$ to be occupied. For our
Slater determinant the constrained minimization is given by the
equation
\begin{equation} 
\delta\int d^2x_\perp {dx^-\over 2}\,
\langle \alpha\mid\Lambda_+
\left({\cal H}_{LF}-{p^-_\alpha\over 2}
\right)\Lambda_+
\mid \alpha
\rangle =0, \label{min} \end{equation}
where the quantities $p^-_\alpha$ are the Lagrange multiplication
factors for each occupied orbital. The relation (\ref{min}) leads
immediately to our mode equation
\begin{eqnarray} 
\lefteqn{p^-_\alpha\Lambda_+\mid \alpha\rangle
=\left(i\partial^++
2g_v\bar{V}^-\right)\Lambda_-\mid \alpha \rangle}\quad&&\nonumber\\
&& +\left(\bbox{\alpha}_\perp\cdot 
(\bbox{p}_\perp-g_v\bbox{\bar V}_\perp)+\beta(M+g_s\phi)\right)
{1\over i\partial^+}
\left(\bbox{\alpha}_\perp\cdot 
(\bbox{p}_\perp-g_v\bbox{\bar V}_\perp)+\beta(M+g_s\phi)\right)
\Lambda_+\mid \alpha \rangle.\label{eign} 
\end{eqnarray}
The operators $\bbox{\alpha}$ and $\beta$ have non-zero values when
appearing between $\Lambda_+$ and $\Lambda_-$, but vanish when
appearing between the two identical projection operators. Thus we may
obtain $\Lambda_- \mid \alpha \rangle$ as
\begin{equation} 
\Lambda_-\mid \alpha\rangle=
{1\over i\partial^+}\left(\bbox{\alpha}_\perp\cdot 
(\bbox{p}_\perp-g_v\bbox{\bar V}_\perp)+\beta(M+g_s\phi)\right)\Lambda_+\mid
\alpha \rangle,
\end{equation}
or
\begin{equation} 
i\partial^+\Lambda_-\mid \alpha\rangle= \left(\bbox{\alpha}_\perp\cdot 
(\bbox{p}_\perp-g_v\bbox{\bar V}_\perp)+\beta(M+g_s\phi)\right)
\Lambda_+\mid\alpha \rangle.
\label{nminus}
\end{equation}
One may use Eq.~(\ref{nminus}) to rewrite Eq.~(\ref{eign}) as
\begin{eqnarray} 
p^-_\alpha\Lambda_+\mid \alpha\rangle&=&\left(i\partial^++
2 g_v\bar{V}^-\right)
\Lambda_+\mid \alpha\rangle\nonumber\\
&&+
\left(\bbox{\alpha}_\perp\cdot 
(\bbox{p}_\perp-g_v\bbox{\bar V}_\perp)+\beta(M+g_s\phi)\right)
\Lambda_-\mid \alpha\rangle\;.
\label{nplus}
\end{eqnarray} 
Equations (\ref{nplus}) and (\ref{nminus}) are the essential results of
this section. We have obtained the light-front version of the Hartree
equations.

\subsection{Nuclear energy}

There are contributions to the expectation value of $P^-+P^+$ from the
nucleons, scalar mesons, and vector mesons. The nucleonic term is given
from the expectation value of the nucleonic part $H_{LF}$~(\ref{hlf}).
Taking the nuclear expectation value of $H_{LF}$ leads to a sum of
matrix elements in the occupied states $\mid \alpha\rangle$. The use of
the wave equation (\ref{eign}) leads to the result 
\begin{equation}
\langle\Psi\mid H_{LF}\mid\Psi\rangle=\sum_\alpha^{\text{occ}}
{p^-_\alpha\over 2},\label{en}
\end{equation}
in which the sum is over $\alpha$ includes only occupied states.
%An
%interaction term $V_I$ arises from summing the matrix elements
%$2\langle \alpha\mid\Lambda^+ g_v V^-\Lambda^+\mid \alpha\rangle$
%arising from the second term on the right hand side of
%Eq.~(\ref{eign}). More explicitly we have 
%\begin{equation}
%V_I=g_v\int d^3x\,\bar{V}^-(\bbox{x})
%\langle\Psi\mid \psi^\dagger(\bbox{x})\psi(\bbox{x})\mid\Psi\rangle.
%\label{vi}
%\end{equation}

The contribution from the scalar mesons $E_s$ is given from the scalar
meson terms of Eqs.~(\ref{pmph}) and (\ref{ppph}) by
\begin{equation}
E_s={1\over2}\int d^2k_\perp dk^+\theta(k^+)\,\left[{k_\perp^2+m_s^2\over
k^+}+k^+\right] \langle\Psi\mid a^\dagger(\bbox{k}) a(\bbox{k})
\mid\Psi\rangle.
\end{equation}
In our mean field approximation
\begin{equation}
\langle\Psi\mid a^\dagger(\bbox{k}) a(\bbox{k})
\mid\Psi\rangle=
|\langle\Psi\mid a(\bbox{k})\mid\Psi\rangle\mid^2,\label{sme}
\end{equation}
with the matrix element already known from Eq.~(\ref{pap}). Then
straightforward calculation leads to the result
\begin{equation}
E_s={1\over2}\int {d^3k\over (2\pi)^3}\,{1\over \bbox{k}^2+m_s^2}
\mid \langle\Psi\mid J(\bbox{k})\mid\Psi\rangle\mid^2,\label{ephi}
\end{equation}
where $J(\bbox{k})$ is given by Eq.~(\ref{jk}), the replacement
(\ref{relate}) is used, and as above $k^+$ is replaced by $k^3$. The
above expression is strikingly familiar --- it is the result obtained
in standard equal-time calculations.

The vector meson contribution to the energy $E_v$ is defined as the
one-half of the sum of the terms of $P_v^\pm$ of Eqs.~(\ref{pvm}) and
(\ref{ppm}). The calculation of $P_v^\pm$ is rather similar to the one
just done for the scalar mesons. One uses the results (\ref{jwk}),
(\ref{pvp}), and $X^\mu$(\ref{xmu}). The effects of the instantaneous
term $v_3$ are cancelled by the non-$\gamma^\mu$ term of $X^\mu$, so
that we find
\begin{equation}
E_v=-{1\over2}\int {d^3k\over (2\pi)^3}\,{1\over \bbox{k}^2+m_v^2}
\mid J_v(\bbox{k})\mid^2,\label{e1v}
\end{equation}
where
\begin{equation}
J_v(\bbox{k})\equiv g_v\langle\Psi\mid\int d^3x\,
e^{i\bbox{k}\cdot\bbox{x}}\psi^\dagger(\bbox{x})\psi(\bbox{x})
\mid\Psi\rangle.
\end{equation}
% in which the spherical nature of the nucleus has been used.

The nuclear mass $M_A$ is then given by
\begin{equation}
M_A=\sum_\alpha^{\text{occ}}{p^-_\alpha\over2} +E_s+E_v,
\end{equation}
with expressions for each of the contributions given above.

\subsection{Relation with the equal-time formulation}

Our main results obtained using the mean field approximation and
including the recoil of the $A-1$ nuclear system are embodied in the
equations (\ref{nplus}) and (\ref{nminus}). We solve these equations
below using a mixed momentum-coordinate space procedure in which the
wave functions are $\langle
p^+,\bbox{x}_\perp\mid\alpha\rangle=\psi_\alpha(p^+,\bbox{x}_\perp)$.
The values of $p^+$ are greater than zero. Thus the so-called spectrum
condition that positive energy particles have only positive
plus-momenta is maintained in our mean field approximation.

An intermediate step is to make an approximation by using coordinate
space techniques. Here one does not maintain the spectrum condition in
an exact manner. Then one can show there is a very close relationship
(but approximate) between our $\psi_\alpha(x^-,\bbox{x}_\perp)$ and the
usual solutions to the Dirac equation obtained from the equal-time ET
formulation.

To see this, let's first consider the case where there is no vector
potential at all ($\bar{V}^\mu\to0$). Then multiply Eq.~(\ref{nminus})
by $\gamma^+$ and Eq.~(\ref{nplus}) by $\gamma^-$. Use
$\gamma^\pm\Lambda_\mp\mid\alpha\rangle
=\gamma^\pm(\Lambda_++\Lambda_-)\mid\alpha\rangle =\gamma^\pm
\mid\alpha\rangle$, and then add the two equations. This gives
\begin{equation}
\left(\gamma^0 p^-_\alpha-\gamma^3(2p^+-p^-_\alpha)\right)
\psi_{\alpha}(x^-,\bbox{x}_\perp)=
2\left(\bbox{\gamma}_\perp\cdot
\bbox{p}_\perp+M+g_s\phi(x^-,\bbox{x}_\perp)\right)
\psi_{\alpha}(x^-,\bbox{x}_\perp).\label{almost}
\end{equation}
Convert this to ordinary coordinates using $x^-=- 2z$, so that
$p^+=i\partial^+= 2i{\partial\over \partial x^-}\to-i{\partial\over
\partial z}$. The operator $p^+$ acts as a $p^3$ operator, and the
result (\ref{almost}) looks like the Dirac equation of the equal-time
formulation, except for the offending term $-p^-_\alpha$ multiplying
the $\gamma^3$. This motivates us to look for a solution of the form
$\psi_\alpha(z,\bbox{x}_\perp)=f(z)\psi^{\text{ET}}_\alpha(z,\bbox{x}_\perp)$,
in which $f(z)$ is chosen to remove to the offending term. The notation
ET refers to the usual equal-time solution, because we see that
$\psi^{\text{ET}}_\alpha$ obeys the usual ET Dirac equation
\begin{equation}
\left(\gamma^0
{p^-_\alpha\over 2}-\bbox{\gamma}\cdot\bbox{p} 
-M-g_s\phi(z,\bbox{x}_\perp)\right)
\psi^{\text{ET}}_\alpha(z,\bbox{x}_\perp)=0,
\end{equation}
provided
\begin{equation}
f(z)=e^{i p^-_\alpha z/2},
\end{equation}
so that
\begin{equation}
\psi_\alpha(z,\bbox{x}_\perp)=e^{i p^-_\alpha z/2}\,
\psi^{\text{ET}}_\alpha(z,\bbox{x}_\perp).
\end{equation}

The quantity of interest is $\psi_\alpha(p^+,\bbox{x}_\perp)$ which is
expressed as 
\begin{equation}
\psi_\alpha(z,\bbox{x}_\perp)\approx{1\over\sqrt{2\pi}}
 \int_0^\infty dp^+\,e^{ip^+z}\psi_\alpha(p^+,\bbox{x}_\perp). \label{inex}
\end{equation}
The approximation is that the correct version of
$\psi_\alpha(p^+,\bbox{x}_\perp)$ will have no support for $p^+<0$, but
the approximation (\ref{inex}) does. We can determine this support by
examining the inverse Fourier transform. This gives
\begin{equation}
\psi_\alpha(p^+,\bbox{x}_\perp)={1\over\sqrt{2\pi}}
\int_{-\infty}^\infty dz\, e^{-i(p^+-p^-_\alpha/2)z}
\psi^{\text{ET}}_\alpha(z,\bbox{x}_\perp),
\label{emcr}
\end{equation}
which is a Fourier transform of the equal-time Dirac wave function at a
$z$-component of momentum $p^+ -p^-_\alpha/2$. This is not exactly
equal to zero when $p^+$ is zero or negative, but it is very small
because $p^-_\alpha/2$ includes the nucleon mass. The relationship
between the term $p^-_\alpha/2$ and the binding energy of the level
denoted by $\alpha$ is
\begin{equation}
p^-_\alpha/2=M-\varepsilon_\alpha.
\end{equation}
Thus the relation (\ref{emcr}) is just the usual equal-time procedure
equal-time prescription, represented by Eq.~(\ref{UP}), of replacing
the kinematic variable $p^+$ by the combination of dynamical and
kinematic variables $M-\varepsilon_\alpha^++p^3$ for the orbital
$\alpha$:
\begin{equation}
p^+\to M-\varepsilon_\alpha^+ + p^3.\label{prescribe}
\end{equation}

However, the prescription (\ref{prescribe}) is dramatically changed
when the vector potential is included. To see this, multiply
Eq.~(\ref{nplus}) by $\gamma^-$ and Eq.~(\ref{nminus}) by $\gamma^+$.
Then Eq.~(\ref{almost}) becomes
\begin{eqnarray}
\lefteqn{
\left(\gamma^0(p^-_\alpha-2g_vV^0)-
\gamma^3(2p^+-p^-_\alpha+2g_vV^0)\right)\psi_{\alpha}(x^-,\bbox{x}_\perp)=}
\qquad\qquad\qquad&&\nonumber\\
&&\qquad\qquad\qquad
2\left( \bbox{\gamma}_\perp\cdot \bbox{p}_\perp+M+g_s\phi
\right)\psi_{\alpha}(x^-,\bbox{x}_\perp),\label{nalmost}
\end{eqnarray}
in which we used $\bar{V}^-=V^-=V^0$. We again wish to reduce the
coefficient of the $\gamma^3$ term to $2p^+$. This can be done with a
new version of the multiplier $f(z)$. We find that the light-front wave
function is given by 
\begin{equation}
\psi_\alpha(p^+,\bbox{x}_\perp)={1\over\sqrt{2\pi}}\int_{-\infty}^\infty
dz\,
e^{-i(p^+-p_\alpha^-/2)z}\,e^{-ig_v\Lambda(z,\bbox{x}_\perp)}\,
\psi^{\text{ET}}_\alpha(z,\bbox{x}_\perp),\label{emcv}
\end{equation}
where 
\begin{eqnarray}
\partial^+\Lambda(x^-,\bbox{x}_\perp)&=&V^0(x^-,\bbox{x}_\perp),\nonumber\\
\Lambda(z,\bbox{x}_\perp)&=&\int_z^\infty dz'\,V^0(z',\bbox{x}_\perp),
\end{eqnarray}
and
\begin{equation}
\gamma^0 (p^-_\alpha-g_vV^0)\psi^{\text{ET}}_\alpha(z,\bbox{x}_\perp)
=(\bbox{\gamma}\cdot \bbox{p}
+M+g_s\phi)\psi^{\text{ET}}_\alpha(z,\bbox{x}_\perp).
\end{equation}
The relation (\ref{emcv}) tells us that the influence of the vector
potential is to remove plus-momentum from the nucleons. This removal
and enhancement of the nuclear vector meson content is the most
dramatic result we have.

How accurate is Eq.~(\ref{emcv})? This can only be addressed by solving
the problem in a manner which respects the spectrum condition. The
results show an astonishing agreement between the eigenvalues of
Eq.~(\ref{nplus}) and those of the equal-time Dirac equation. Thus it
should be safe to use Eq.~(\ref{emcv}) for qualitative purposes.

\section{Technical Aspects}

The solution of the nucleon and meson field equations are discussed. The
reduction of Eqs.~(\ref{nplus}) and (\ref{nminus}) to a two-dimensional
matrix equation is presented here. The new numerical technique 
involving splines is elaborated in App.~\ref{appa}.

The nucleon mode equation resulting from the minimization of
$\case1/2(P^+ + P^-)$ is given by the coupled set of equations
(\ref{nplus}) and (\ref{nminus}). The meson fields $\phi$ and $V^\pm$
obey the equations
\begin{eqnarray}
\left(-(\partial^+)^2-\partial_\perp^2 + m_s^2 \right)
\phi(x^-,\bbox{x}_\perp) &=& -g_s
\sum_\alpha^{\text{occ}} \bar\psi_\alpha(x^-,\bbox{x}_\perp)
\psi_\alpha(x^-,\bbox{x}_\perp),\\
\left(-(\partial^+)^2-\partial_\perp^2 + m_v^2 \right)
V^\pm(x^-,\bbox{x}_\perp) &=& g_v
\sum_\alpha^{\text{occ}} \bar\psi_\alpha(x^-,\bbox{x}_\perp)
\gamma^\pm \psi_\alpha(x^-,\bbox{x}_\perp),
\end{eqnarray}
in which $\psi_\alpha(x^-,\bbox{x}_\perp)\equiv
\langle x^-,\bbox{x}_\perp\mid \alpha\rangle$.
We use the Harindranath-Zhang \cite{harizhang} representation for the
Dirac matrices $\alpha$ and $\beta$, which allows us to write
Eqs.~(\ref{nplus}) and (\ref{nminus}) in 2-component form. This
representation can be obtained from the standard representation
\begin{equation}
\alpha_i = \left(\begin{array}{cc}
0 & \sigma_i\\ \sigma_i & 0 \end{array}\right), \qquad
\beta = \left(\begin{array}{cc}
1 & 0 \\ 0 & -1 \end{array}\right),
\end{equation}
by the unitary transformation
\begin{equation}
U = {1\over \sqrt2} \left(\begin{array}{cc}
1&-\sigma_3\\ \sigma_3&1 \end{array} \right).
\end{equation}
Hence $\psi\to U \psi$ and $\theta \to U \theta
U^\dagger$, where $\theta$ is a Dirac matrix in the standard
representation. In our representation, the matrices of interest are:
\begin{eqnarray}
\Lambda^+ = \left(\begin{array}{cc} 0&0\\0&1 \end{array}\right),\qquad&&
\Lambda^- = \left(\begin{array}{cc} 1&0\\0&0 \end{array}\right),\nonumber\\
\beta = \left(\begin{array}{cc}0&\sigma_3\\ \sigma_3&0
\end{array}\right),\qquad&&
\alpha_3 = \left(\begin{array}{cc} -1&0\\ 0&1 \end{array}\right),\qquad
\alpha_\perp = \left(\begin{array}{cc} 0&\sigma_\perp\\ \sigma_\perp& 0
\end{array}\right).
\end{eqnarray}

The 4-component wavefunction $|\psi_\alpha\rangle$ may now be written
in the form
\begin{equation}
\langle x^-,\bbox{x}_\perp| \psi_\alpha\rangle =
\left(\begin{array}{rr}\langle x^-,\bbox{x}_\perp| \psi_\alpha^-\rangle \\
\langle x^-,\bbox{x}_\perp|\psi_\alpha^+\rangle \end{array}\right),
\end{equation}
in terms of the 2-component wavefunctions $|\psi_\alpha^+\rangle$ and
$|\psi_\alpha^-\rangle$. Thus the 2-component form of Eqs.~(\ref{nplus})
and (\ref{nminus}) is
\begin{mathletters}
\begin{eqnarray}
(p_\alpha^- - 2g_v V^- -i\partial^+) |\psi_\alpha^+\rangle
&=& \left(\bbox{\sigma}_\perp\cdot (\bbox{p}_\perp -
g_v \bbox{\partial}_\perp\Lambda) +
\sigma_3 (M + g_s \phi) \right) |\psi_\alpha^-\rangle,\\
i\partial^+ |\psi_\alpha^-\rangle
&=& \left(\bbox{\sigma}_\perp\cdot (\bbox{p}_\perp -
g_v \bbox{\partial}_\perp\Lambda) +
\sigma_3 (M + g_s \phi) \right) |\psi_\alpha^+\rangle.
\end{eqnarray}
\end{mathletters}

The scalar and vector densities are defined as
\begin{eqnarray}
\rho^s &\equiv& \sum_\alpha^{\text{occ}}\bar\psi_\alpha
\psi_\alpha
= \sum_\alpha^{\text{occ}}\left(\bar\psi_\alpha^+
\sigma_3 \psi_\alpha^- + \bar\psi_\alpha^- \sigma_3 \psi_\alpha^+\right),
\label{rhosdef}\\
\rho^\pm &\equiv& \sum_\alpha^{\text{occ}}
\bar\psi_\alpha \gamma^\pm \psi_\alpha =
\sum_\alpha^{\text{occ}} 2(\psi_\alpha^\pm)^\dagger \psi_\alpha^\pm\,.
\label{rhodef}
\end{eqnarray}
In the nuclear rest frame, $\rho^+ = \rho^- = \rho^0$, where $\rho^0$
is the usual nucleon density. Hence $V_\perp=0$ and $V^-=V^+$ in this
frame.

\subsection{Angular momentum}

We can write
\begin{equation}
\bbox{\sigma}_\perp \cdot \bbox{p}_\perp = \sigma_{(+)} p_{(-)} + \sigma_{(-)}
p_{(+)},
\end{equation}
where $\sigma_{(\pm)} = \frac{1}{2} (\sigma_1 \pm i \sigma_2)$ and
\begin{equation}
p_{(\pm)} = p_1 \pm i p_2 = - i e^{\pm i \phi}
\left({\partial\over \partial r} \pm {1\over r}
{\partial\over \partial\phi}\right).
\end{equation}
Here $r = |\bbox{x}_\perp|$ and $\phi$ is the azimuthal angle, using
cylindrical coordinates. For the
nuclear physics problems of interest, we anticipate that there is an
axis of azimuthal symmetry. Hence we can expand the 2-component
wavefunctions in eigenstates of angular momentum $J_z$, with eigenvalue
$j_z$:
\begin{eqnarray}
\langle x^-,\bbox{x}_\perp|\psi_\alpha^\pm\rangle &=&
 i \langle x^-,r|u_\alpha^\pm\rangle
e^{i (j_z - \case1/2)\phi}\chi_{\case1/2}
+ \langle x^-,r|l_\alpha^\pm\rangle
e^{i (j_z + \case1/2)\phi}\chi_{-\case1/2}\nonumber\\
&=& \left(\begin{array}{r} i u_\alpha^\pm(x^-,r) e^{i (j_z - \case1/2)\phi}\\
l_\alpha^\pm(x^-,r) e^{i (j_z + \case1/2)\phi}\end{array}\right),
\end{eqnarray}
where $\chi_{\case1/2}$ and $\chi_{-\case1/2}$ are the 2-component Pauli
spinors.

The equations to be solved are then
\begin{mathletters}
\label{ameqns}
\begin{eqnarray}
\left(p_\alpha^- - 2g_v V^+ - i\partial^+ \right) u_\alpha^+ &=&
- \left({\partial\over \partial r} + {j_z+\case1/2\over r}
- i g_v{\partial\Lambda\over \partial r}\right)
l_\alpha^- + M^* u_\alpha^-,\\
\left(p_\alpha^- - 2g_v V^+ - i\partial^+ \right) l_\alpha^+ &=&
\left({\partial\over \partial r }- {j_z-\case1/2\over r}
- i g_v{\partial\Lambda\over \partial r}\right) u_\alpha^-
- M^* l_\alpha^-,\\
i\partial^+ u_\alpha^- &=&
- \left({\partial\over \partial r} + {j_z+\case1/2\over r}
- i g_v{\partial\Lambda\over \partial r}\right) l_\alpha^+
+ M^* u_\alpha^+,\\
i\partial^+ l_\alpha^- &=&
\left({\partial\over \partial r} - {j_z-\case1/2\over r}
- i g_v{\partial\Lambda\over \partial r}\right) u_\alpha^+
- M^* l_\alpha^+. 
\end{eqnarray}
\end{mathletters}
The wavefunctions $u_\alpha^\pm$ and $l_\alpha^\pm$, the nucleon
effective mass $M^*=M+g_s \phi$, the vector potential $V^+$, and
$\Lambda$ are all functions of both $x^-$ and $r$. Eqs.~(\ref{ameqns})
have a manifest spin degeneracy under $j_z \to -j_z$. 
Solutions with the same eigenvalue $p_\alpha^-$ are obtained with the
corresponding replacement
\begin{equation}
\left(\begin{array}{l}
u_\alpha^+\\l_\alpha^+\end{array}\right) \to
\left(\begin{array}{l}
l_\alpha^+\\u_\alpha^+\end{array}\right), \qquad
\left(\begin{array}{l}
u_\alpha^-\\l_\alpha^-\end{array}\right) \to
- \left(\begin{array}{l}
l_\alpha^-\\u_\alpha^-\end{array}\right),
\end{equation}
Combined with isospin symmetry, we therefore have a manifest fourfold
degeneracy of each single particle state. The numerical solution to
Eqs.~(\ref{ameqns}) is discussed in App.~\ref{appa}.

\section{Nuclear Binding Energies}

If these solution to Eqs.~(\ref{nplus}) and (\ref{nminus}) are to have
any relevance at all, they should respect rotational invariance. The
success in achieving this is examined in Tables I and II, which give
our results for the spectra of $^{16}$O and $^{40}$Ca, respectively.
Scalar and vector meson parameters are taken from Horowitz and
Serot~ \cite{hs}, and we have assumed isospin symmetry. We see that the
$J_z=\pm1/2$ spectrum contains the eigenvalues of all states, since all
states must have a $J_z=\pm1/2 $ component. Furthermore, the essential
feature that the expected degeneracies among states with different
values of $J_z$ are reproduced numerically. 

\begin{table}[t]
\caption{Comparison of the single particle spectra of $^{16}$O in the
equal-time (ET) formalism ($E_\alpha-M$) with the light-front
(LF) method ($p_\alpha^-/2-M$). Units are in MeV.}
\rule{0in}{2ex}
\begin{center}
\begin{tabular}{lddd}
\multicolumn{2}{c}{ET} & \multicolumn{2}{c}{LF} \\
\cline{1-2} \cline{3-4}
State $\alpha$ & $E_\alpha-M$ & $J_z=\pm 1/2$ & $J_z=\pm 3/2$\\
\cline{1-1} \cline{2-2} \cline{3-3} \cline{4-4} 
0s$_{1/2}$ & $-$41.73 & $-$41.73 & \\
0p$_{3/2}$ & $-$20.77 & $-$20.79 & $-$20.77 \\
0p$_{1/2}$ & $-$12.49 & $-$12.51 & \\
\end{tabular}
\end{center}
\label{table1}
\end{table}

\begin{table}
\caption{Comparison of the ET and LF single particle spectra of
$^{40}$Ca.}
\rule{0in}{2ex}
\begin{center}
\begin{tabular}{ldddd}
\multicolumn{2}{c}{ET} & \multicolumn{3}{c}{LF} \\
\cline{1-2} \cline{3-5}
State $\alpha$ & $E_\alpha-M$ & $J_z=\pm 1/2$ & $J_z=\pm 3/2$
& $J_z=\pm 5/2$\\
\cline{1-1} \cline{2-2} \cline{3-3} \cline{4-4} \cline{5-5}
0s$_{1/2}$ & $-$55.40 & $-$55.39 & & \\
0p$_{3/2}$ & $-$38.90 & $-$38.91 & $-$38.90 & \\
0p$_{1/2}$ & $-$33.18 & $-$33.18 & & \\
0d$_{5/2}$ & $-$22.75 & $-$22.76 & $-$22.75 & $-$22.74 \\
1s$_{1/2}$ & $-$14.39 & $-$14.36 & & \\
0d$_{3/2}$ & $-$13.87 & $-$13.88 & $-$13.89 & \\
\end{tabular}
\end{center}
\label{table2}
\end{table}

\begin{table}
\caption{Total plus-momentum per nucleon for $^{16}$O, $^{40}$Ca,
$^{80}$Zr, and nuclear matter (NM) in MeV. No Coulomb interaction is
included here.}
\rule{0in}{2ex}
\begin{center}
\begin{tabular}{ldddd}
Nucleus & $P_N^+/A$ & $P_s^+/A$ & $P_v^+/A$ & $P^+/A$\\
\hline
$^{16}$O & 704.7 & 6.4 & 221.8 & 932.9\\
$^{40}$Ca & 672.6 & 4.7 & 253.3 & 930.6\\
$^{80}$Zr & 655.2 & 3.6 & 270.2 & 929.0\\
NM & 569.0 & 0.0 & 354.2 & 923.2\\
\end{tabular}
\end{center}
\label{table3}
\end{table}

The results shown in Tables I--III are obtained using a basis of 20
splines, a box size of $2L = 24$ fm, and 24 Fourier components in the
expansion of the wavefunction (see App.~\ref{appa}). This
value of $L$ is large enough so that our results do not depend on it,
and the number of terms in the expression for the density is enough to
ensure that the densities are spherically symmetric. Another feature is
that the spectrum with $p^+ > 0$ has no negative energy states, so
that in using the LF method one is working in a basis of positive
energy states only.

The values of $p_\alpha^-/2$ given in Tables I and II are essentially
the same as the single particle energies $E_\alpha$ of the ET
formalism, to within the expected numerical accuracy of our program.
This equality is not mandated by spherical symmetry alone because the
solutions in the equal-time framework have non-vanishing components
with negative values of $p^+$. 

Table III gives the contributions to the total $P^+$ momentum from the
nucleons, scalar mesons, and vector mesons for $^{16}$O, $^{40}$Ca, and
$^{80}$Zr, as well as the nuclear matter limit. In the next section we
examine in detail the momentum distributions giving rise to these
expectation values.

\section{Plus-Momentum Distributions and Lepton-Nucleus Deep Inelastic
Scattering }

We discuss the probability that a nucleon, or meson has a momentum
$p^+$. In the light-front formulation, these distribution functions are
determined by the absolute square of the ground state wave function.
Each distribution is discussed in turn.

\subsection{Nucleon plus-momentum distribution}

The light-front formulation is very useful for obtaining this
observable. The probability that we want, $f_N(p^+)$, follows from
Eq.~(\ref{rhodef}) as
\begin{equation}
f_N(p^+) = 2 \sum_\alpha^{\text{occ}} \int d^2 x_\perp\,
\mid\langle p^+,\bbox{x}_\perp \mid
\psi_\alpha^+\rangle\mid^2,\label{fnp}
\end{equation}
with
\begin{eqnarray}
A &=& \int_0^\infty dp^+\, f_N(p^+),\label{fnnorm}\\
P_N^+ &=& \int_0^\infty dp^+\, p^+ f_N(p^+).\label{fnpn}
\end{eqnarray}
The next step is to define a dimensionless variable $y$:
\begin{equation}
y\equiv p^+{ A\over M_A} \equiv {p^+\over \bar M_A},
\end{equation}
and a dimensionless distribution $f_N(y)$:
\begin{equation}
f_N(y)\equiv{ f_N(p^+)\over \bar M_A}.
\end{equation}

\begin{figure}
%\begin{Large}
\unitlength1.cm
\begin{picture}(15,9)(-14,1) %15,9)(2,1)
\includegraphics{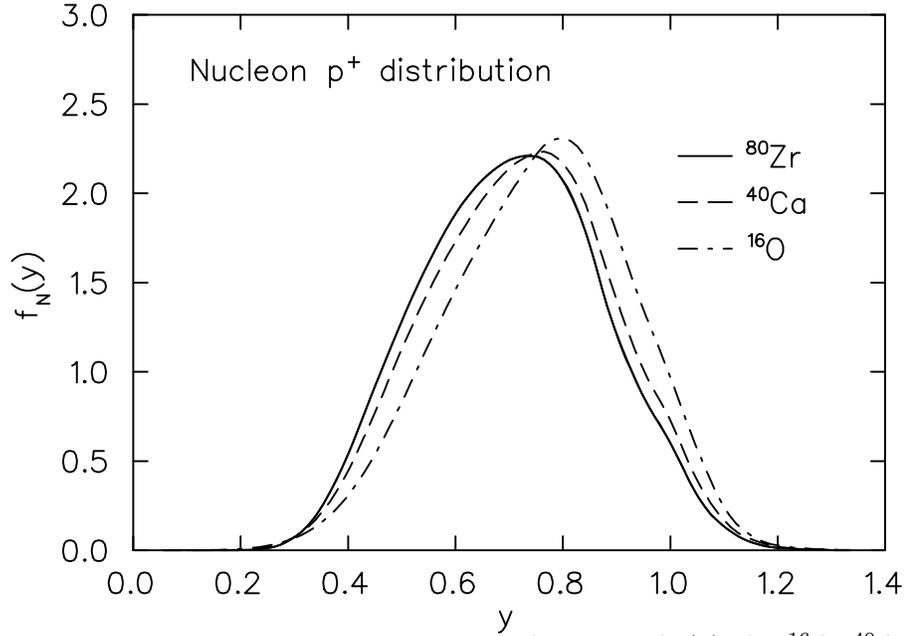}
\end{picture}
%\end{Large}
\caption{Nucleon plus-momentum distribution function, $f_N(y)$, for
$^{16}$O, $^{40}$Ca, and $^{80}$Zr. Here $y\equiv p^+/(M_A/A)$.}
\label{fig:fny}
\end{figure}

The result is shown in Fig.~\ref{fig:fny} for $^{16}$O, $^{40}$Ca, and
$^{80}$Zr. The peaks of the distributions range from $y\approx 0.72$ to
$y\approx 0.80$, whereas the average values $\langle y\rangle$ are
somewhat lower (see Table III). The distribution is not symmetric about
its average value, as it would be if a simple Fermi gas model were
used. Both of these effects are caused by the presence of nuclear
mesons, which carry the remainder of the plus-momentum.

\begin{figure}
\unitlength1.cm
\begin{picture}(15,9)(-14,1)
\includegraphics{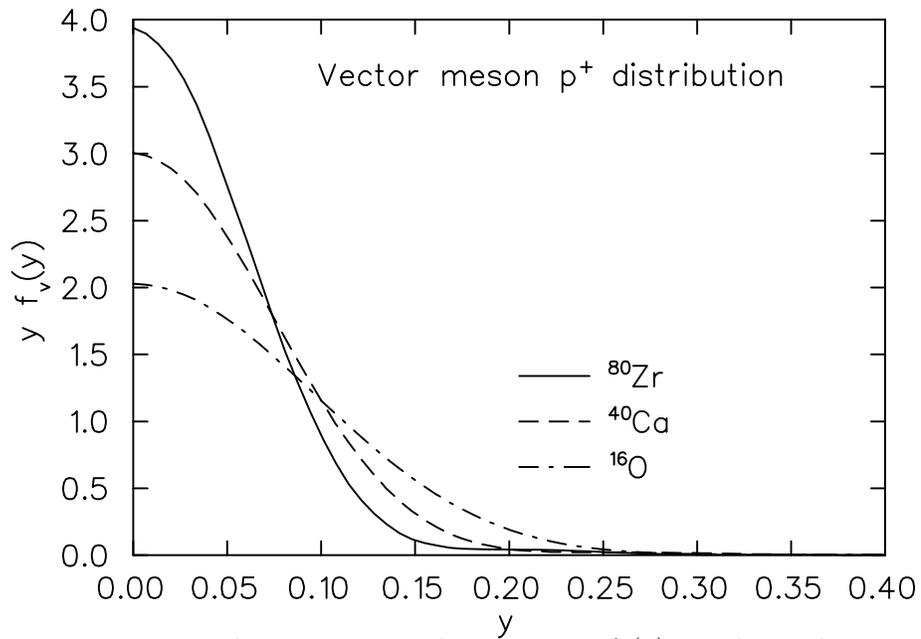}
\end{picture}
\caption{Vector meson plus-momentum distribution $y f_v(y)$. In the
nuclear matter limit, $y f_v(y)$ becomes a delta function.}
\label{fig:fvy}
\end{figure} 

\subsection{Scalar meson distribution }

The probability we want is given by 
\begin{equation}
f_s(k^+)=\int d^2k_\perp \,\langle\Psi\mid a^\dagger(\bbox{k})
a(\bbox{k})\mid\Psi\rangle.
\end{equation} 
Using Eqs.~(\ref{sme}) and (\ref{pap}) this becomes
\begin{equation}
f_s(k^+)=\int { d^2k_\perp \over (2\pi)^3}\,{2k^+\over
\left(\bbox{k}^2+m_s^2\right)^2}
\mid \langle\Psi\mid J(\bbox{k})\mid\Psi\rangle\mid^2.\label{f1phi}
\end{equation}
This result which is of the same form as in Ref.~\cite{bm98}. 
A final step is to define a dimensionless distribution $f_s(y)$:
\begin{equation}
f_s(y)\equiv{ f_s(k^+)\over \bar M_A}.
\end{equation}

The scalar mesons are found to carry less that 1\% of the plus-momentum of
the nucleus (Table III), which is negligible. 

\subsection{Vector meson distribution}

The probability we want is given by 
\begin{equation}
f_v(k^+)=\int d^2k_\perp\, \sum_{\omega=1,3}
\langle\Psi\mid a^\dagger(\bbox{k},\omega)
a(\bbox{k},\omega)\mid\Psi\rangle.
\end{equation} 
Using Eqs.~(\ref{pvp}) and the mean field approximation
(\ref{replace1}) this becomes
\begin{equation}
f_v(k^+)=\int { d^2k_\perp \over (2\pi)^3}\,
{2k^+\over\left( \bbox{k}^2+m_v^2\right)^2}
\sum_{\omega=1,3}
\mid J(\bbox{k},\omega)\mid^2,\label{f1v}
\end{equation}
in which
\begin{equation}
J(\bbox{k},\omega)=\int d^3x\, e^{i\bbox{k}\cdot\bbox{x}}
\langle\Psi\mid \bar \psi(\bbox{x})\gamma\psi(\bbox{x})\cdot \bar\epsilon
(\bbox{k},\omega)\mid\Psi\rangle.
\end{equation}
Using Eq.~(\ref{ebar}) and
that only the $\mu=\nu=-$ term enters, leads to the
result that 
\begin{equation}
f_v(k^+)=\int { d^2k_\perp \over (2\pi)^3}\,{2k^+\over
\left( \bbox{k}^2+m_v^2\right)^2}
{k_\perp^2+m_v^2\over {k^+}^2}\mid J_v(\bbox{k})\mid^2,\label{f2v}
\end{equation}
another result which is of the same form as in Ref.~\cite{bm98}.
A final step is to define a dimensionless distribution $f_v(y)$
\begin{equation}
f_v(y)\equiv{ f_v(k^+)\over \bar M}.
\end{equation}

The vector mesons carry approximately 30\% of the nuclear
plus-momentum. The technical reason for the difference with the scalar
mesons is that the evaluation of
$a^\dagger(\bbox{k},\omega)a(\bbox{k},\omega)$ counts vector mesons
``in the air''and the resulting expression contains polarization
vectors that give a factor of ${1\over k^+}$ in Eq.~(\ref{f2v}) which
enhances the distribution of vector mesons of low $k^+$. The results
for the vector meson distribution are shown in Fig.~\ref{fig:fvy}. 
Clearly as the size of the nucleus increases the enhancement of the
distribution at lower values of $k^+$ becomes more evident. In the case
of nuclear matter the distribution $k^+ f_v(k^+)$ becomes a delta
function.

\subsection{Lepton-nucleus deep inelastic scattering}

It is worthwhile to see how the present results are related to 
lepton-nucleus deep inelastic scattering experiments. We find that the
nucleons carry only about 70\% of the plus-momentum. The use of our
$f_N$ in standard convolution formulae lead to a reduction in the
nuclear structure function that is far too large ($\sim$95\% is
needed \cite{emcrevs}) to account for the reduction
observed \cite{emcrevs} in the vicinity of $x\sim 0.5$. The reason for
this is that the quantity $M +g_s\phi$ acts as a nucleon effective mass
of about 670 MeV, which is very small. A similar difficulty occurs in
the $(e,e')$ reaction \cite{frank} when the mean field theory is used
for the initial and final states. The use of a small effective mass and
a large vector potential enables a simple reproduction of the nuclear
spin orbit force \cite{bsjdw,hs}. However, effects beyond the mean field
may lead to a significant effective tensor coupling of the isoscalar
vector meson \cite{ls} and to an increased value of the effective mass.
Such effects are incorporated in Bruckner theory, and a light-front
version \cite{rmgm98} could be applied to finite nuclei with better
success in reproducing the data.

\section{Summary and Discussion}

The previous Sections present a derivation of a light-front version of
mean field theory. The necessary technique is to minimize expectation
value of the sum $P^-+P^+$. This leads to a new set of coupled
equations (\ref{nplus}) and (\ref{nminus}) for the single nucleon
modes. These depend on the meson fields of Eqs.~(\ref{phieq}) and
(\ref{veq}).

The most qualitatively startling feature emerging from the derivation
is that the meson field equations (\ref{phieq}) and (\ref{veq}) are the
same as that of the usual theory, except that $z$ of the equal-time
theory translates to $-x^-/2$ of the light-front version. This can be
understood in a simple manner by noting that light-front quantization
occurs at $x^+=0$. If one then sets $z=-t$, then $x^-=t-z=-2z$.
However, this simple argument is not really justified, because using
$x^\pm$ precludes the use of $z$ and $t$. A general argument, using
the feature that a static source in the usual coordinates corresponds
to a source moving with a constant velocity in light front coordinates,
will be  presented in a separate paper \cite{us2}. That paper also contains a
number of solutions of toy models.

Even though the meson field equations of the light-front and equal-time
theories are the same, there are substantial and significant
differences between the two theories. In our treatment, the mesonic
fields are treated as quantum field operators. The mean field
approximation is developed by replacing these operators by their
expectation values in the complete ground state nuclear wave function.
This means that the ground state wave function contains Fock terms with
mesonic degrees of freedom. We can therefore compute expectation
values other than that of the field. In particular, we are able to
obtain the mesonic momentum distributions (Sec.~VI). This feature has
been absent in standard approaches.

We obtain an approximate solution (\ref{emcv}) of our nucleon mode
equation. Our nucleon mode functions are approximately a phase factor
times the usual equal-time mode functions (evaluated at $x^-=-2z$).
This shows that the energy eigenvalues of the two theories should have
very similar values. But the wave functions are different-- the
presence of the phase factor explicitly shows that the nucleons give up
substantial amounts of plus momentum to the vector mesons.

A new numerical technique, discussed in Sec.~IV and App.~\ref{appa},
is introduced to solve the coupled nucleon and meson field equations.
Our results display the expected $2j_\alpha +1$ degeneracy of the
single nucleons levels, and the resulting binding energies are
essentially the same as for the usual equal-time formulation. This
indicates that the approximation (\ref{emcv}) is valid.

As discussed in Sec.~VI-D, the present results related to 
lepton-nucleus deep inelastic scattering experiments and $(e,e')$
reactions are not consistent with experimental findings. This is
because, in $^{40}$Ca for example, the nucleons carry only 72\% of the
plus momentum. This is a result of the quantity $M +g_s\phi$, which
acts as a nucleon effective mass, is very small, about 670 MeV. The use
of a small effective mass and a large vector potential enables a simple
reproduction of the nuclear spin orbit force \cite{bsjdw,hs}. However,
effects beyond the mean field may lead to a significant effective
tensor coupling of the isoscalar vector meson \cite{ls} and to an
increased value of the effective mass. Such effects are incorporated in
Bruckner theory \cite{rmgm98} which, for infinite nuclear matter,
results in nucleons having about 80-85\% of the nuclear plus-momentum.
A light-front version \cite{rmgm98} should be applied to finite nuclei
with better success in reproducing the data. Another approach could be
to use different Lagrangians, with non-linear couplings between scalar
mesons and the nucleons \cite{bsjdw}, or ones in which the coupling is
of derivative form \cite{zim}: $\bar \psi \gamma^\mu\psi \partial_\mu
\phi$. These models are known to have significantly smaller magnitudes
of the scalar and vector potentials. In particular, in nuclear matter
vector mesons carry only about 10-15\% of the nuclear-plus momentum.
Another interesting possibility would be to obtain a light-front
version of the quark-meson coupling model \cite{qmc}, in which confined
quarks interact by exchanging mesons with quarks in other nucleons.
This model, also has smaller magnitudes of the scalar and vector
potentials.

In any case, these kinds of nuclear physics calculations can be done in
a manner in which modern nuclear dynamics is respected, boost
invariance in the $z$-direction is preserved, and in which the
rotational invariance so necessary to understanding the basic features
of nuclei is maintained.

\acknowledgements

P.G.B. was supported in part by the Natural Sciences and Engineering
Research Council of Canada.

\appendix

\section{Numerical Techniques}
\label{appa}

There are many possible numerical approaches to solving
Eqs.~(\ref{ameqns}). We choose a method that is robust, and emphasizes
the physical content of the wavefunctions, at the expense of being
computationally intensive. We begin by making a Fourier expansion of
the wavefunctions in the variable $x^-$:
\begin{equation}
u_\alpha^\pm(x^-,r) = {1\over \sqrt{2L\,2\pi}} \sum_n e^{-i p_n^+ x^-/2}
u_{\alpha,n}^\pm(r),
\end{equation}
with a similar expansion for $l_\alpha^\pm(x^-,r)$. Boundary conditions
are imposed by constraining the system to be in a ``box'' of a given
length in the variable $x^-$. In the nuclear rest frame, $x^- = - 2 z$,
and so $\partial^+ = 2 \partial_- = -\partial/\partial z$. Hence for
$-L\le z\le +L$, we write
\begin{equation}
\label{decomp}
u_\alpha^\pm(z,r) = {1\over \sqrt{2L\,2\pi}} \sum_n e^{i p_n^+ z}
u_{\alpha,n}^\pm(r),
\end{equation}
with $\{p_n^+ = n q, n=1,2,3,\ldots\}$, and $q = \pi/L$.

For bound states, the functions $u_{\alpha,n}^\pm(r)$ and
$l_{\alpha,n}^\pm(r)$ are real, and
so the scalar and vector densities take the form
\begin{eqnarray}
\rho^s(z,r) &=& \sum_{m\ge 0} \rho_m^s(r) \cos{m q z},\\
\rho^\pm(z,r) &=& \sum_{m\ge 0} \rho_m^\pm(r)
\cos{m q z},\label{densexpand}
\end{eqnarray}
with $m=0,1,2,\ldots$, and
\begin{eqnarray}
\rho_m^s(r) &=& {2\over 2L\,2\pi}{1\over 1+\delta_{m,0}}
\sum_\alpha^{\text{occ}} \sum_n \left[ u_{\alpha,n}^+(r) u_{\alpha,n+m}^-(r)
+ u_{\alpha,n}^-(r) u_{\alpha,n+m}^+(r) \right.\nonumber\\
&&\left.\qquad- l_{\alpha,n}^+(r) l_{\alpha,n+m}^-(r) -
l_{\alpha,n}^-(r) l_{\alpha,n+m}^+(r)\right],\\
\rho_m^\pm(r) &=& {2\over 2L\,2\pi}{2\over 1+\delta_{m,0}}
\sum_\alpha^{\text{occ}} \sum_n \left[u_{\alpha,n}^\pm(r)
u_{\alpha,n+m}^\pm(r) + 
l_{\alpha,n}^\pm(r) l_{\alpha,n+m}^\pm(r)\right].
\end{eqnarray}
The normalization integral for a nucleus with $A$ nucleons is
\begin{eqnarray}
A &=& 2\pi \int_0^\infty dr\,r\,\int_{-L}^L dz\,\rho^+(z,r)\nonumber\\
&=& 2\pi\,2L \int_0^\infty dr\,r\,\rho_0^+(r)\nonumber\\
&=& 2 \int_0^\infty dr\,r\,
\sum_\alpha^{\text{occ}}\sum_n \left[(u_{\alpha,n}^+(r))^2 +
(l_{\alpha,n}^+(r))^2\right].
\end{eqnarray}

\subsection{Meson fields}

The equations for the meson fields are solved using Green function
methods. We illustrate this for the vector field $V^+$, with results
for $\phi$ following by analogy. Starting with
\begin{equation}
\left( -{\partial^2\over \partial z^2} - {\partial^2\over \partial r^2}
-{1\over r} {\partial \over \partial r} + m_v^2\right) V^+(z,r) = g_v
\rho^+(z,r),
\end{equation}
we expand $V^+(z,r)$ in the same form as the density $\rho^+(z,r)$,
Eq.~(\ref{densexpand}):
\begin{equation}
V^+(z,r) = \sum_m V_m^+(r) \cos{m q z}.\label{potexpand}
\end{equation}
The functions $V_m^+(r)$ satisfy
\begin{equation}
\left( - {\partial^2\over \partial r^2}
-{1\over r} {\partial \over \partial r} + m_v^2+ m^2 q^2\right) V_m^+(r) = g_v
\rho_m^+(r),
\end{equation}
and their solution may be written as
\begin{equation}
\label{mesonint}
V_m^+(r) = g_v \int_0^\infty dr'\,r'\,G(r,r') \rho_m^+(r').
\end{equation}
The Green function is
\begin{equation}
G(r,r')=I_0(m_v^* r) K_0(m_v^* r') \theta(r'-r) + 
I_0(m_v^* r) K_0(m_v^* r') \theta(r-r').
\end{equation}
We have introduced the definition $m_v^*\equiv \sqrt{m_v^2 + m^2 q^2}$,
and $I_0$ and $K_0$ are modified cylindrical Bessel functions of zeroth
order. The meson fields are computed numerically from
Eq.~(\ref{mesonint}) by an outward and an inward integration.

\subsection{Solution of nucleon equation}

To streamline the notation, we drop the explicit dependence on the
single particle label $\alpha$ in this section. Equation (\ref{ameqns})
can be rewritten in the form of a $2\times 2$ matrix equation
\begin{equation}
\label{mateqn}
p^- \left(\begin{array}{l} \langle z,r|u^+\rangle \\
\langle z,r|l^+\rangle \end{array}\right)
= \left\{ \left(2 g_v V^+ - i
\textstyle{\partial \over \partial z}\right)I +
{\cal H} {1\over \left(-i{\partial \over
\partial z}\right)I} {\cal H} \right\}
\left(\begin{array}{l} \langle z,r|u^+\rangle \\
\langle z,r|l^+\rangle \end{array}\right),
\end{equation}
with the constrained subsidiary relation
\begin{equation}
\left(\begin{array}{l} \langle z,r|u^-\rangle \\
\langle z,r|l^-\rangle \end{array}\right)
= {1\over \left(-i{\partial \over
\partial z}\right)I} {\cal H} 
\left(\begin{array}{l} \langle z,r|u^+\rangle \\
\langle z,r|l^+\rangle \end{array}\right).
\end{equation}
Here $I$ is the $2\times 2$ identity matrix, and
\begin{eqnarray}
\label{matdefs}
{\cal H} &=& \left(\begin{array}{cc} M^* &
D_1 + i g_v{\partial\Lambda\over \partial r} \\
D_2 - i g_v{\partial\Lambda\over \partial r} &
-M^*\end{array} \right),\\
D_1 &=& -\left({\partial \over \partial r} +
{{j_z + \case1/2}\over r}\right),\\
D_2 &=& \left({\partial \over \partial r} -
{{j_z - \case1/2}\over r}\right).
\end{eqnarray}
If we take $N$ Fourier components $n=1,2,3,\ldots,N$ in the expansion
of Eq.~(\ref{decomp}) in $z$, then $u^+(z,r)=\langle z,r|u^+\rangle$
and $l^+(z,r)=\langle z,r|l^+\rangle$ have the matrix representation
\begin{equation}
\left(\begin{array}{c} u_1^+(r) \\ u_2^+(r) \\ \vdots
\\u_N^+(r) \end{array}\right),\qquad
\left(\begin{array}{c} l_1^+(r) \\ l_2^+(r) \\ \vdots
\\l_N^+(r) \end{array}\right),
\end{equation}
where $u_n^+(r)=\langle p_n^+,r|u^+\rangle$ and
$l_n^+(r)=\langle p_n^+,r|l^+\rangle$.

Equation (\ref{mateqn}) becomes a $2N\times 2N$ matrix equation. Matrix
elements of the $N\times N$ sub-blocks are determined from the
integrals
\begin{eqnarray}
\left[V^+(r)\right]_{(nn')} &=& \langle p_n^+ | V^+(z,r)
|p_{n'}^+\rangle\\
&=& {1\over 2L} \int_{-L}^L
dz\,e^{i(p_{n'}^+-p_n^+)z}\, V^+(z,r),\\
&=&{1+\delta_{m,0}\over 2} V_m^+(r) \delta_{|n-n'|,m}\, .
\end{eqnarray}
Similarly,
\begin{eqnarray}
\left[M^*(r)\right]_{(nn')} &=& M \delta_{n,n'}+ {1+\delta_{m,0}\over 2}
g_s \phi_0(r) \delta_{|n-n'|,m}\, ,\\
\left[-i \textstyle{\partial \over \partial z}\right]_{(nn')} &=& p_n^+
\delta_{n,n'}\, ,\\
\left[D_1\right]_{(nn')} &=& D_1 \delta_{n,n'}\, ,\\
\left[D_2\right]_{(nn')} &=& D_2 \delta_{n,n'}\, ,\\
\left[i \Lambda(r)\right]_{(nn')} &=&
{-\case1/2 V_m^+(r) + (-1)^m V_0^+(r)\over p_n^+ -p_{n'}^+}
\delta_{|n-n'|,m}\quad m\neq 0\, .\label{lambdame}
\end{eqnarray}
The last relation comes from the definition
$-{\partial \Lambda\over \partial z} = V^+$. Using integration by
parts, and a careful treatment of surface terms, gives matrix elements
in the form (\ref{lambdame}).

The problem has now been reduced to an eigenvalue problem involving
$2N$ coupled differential equations in the variable $r$. To solve this,
we make a further expansion of $u_n^\pm(r)$ and $l_n^\pm(r)$ in a
finite basis of B-splines of degree $k$ \cite{deboor,mcneil}:
\begin{eqnarray}
u_n^\pm(r) &=& \sum_{i=1}^{\cal N} \alpha_{n,i}^\pm B_i^{(k)}(r),\\
l_n^\pm(r) &=& \sum_{i=1}^{\cal N} \beta_{n,i}^\pm B_i^{(k)}(r).
\end{eqnarray}
The B-splines $\{B_i^{(k)}, i=1,\ldots,{\cal N}\}$ are polynomials of
degree $k-1$ spanning a domain of equally spaced knots $\{t_i,
i=1,\ldots,{\cal N}+k \}$ in $r$. They are smooth ``local'' functions
that are non-zero only on the interval $t_i < r < t_{i+k}$. This basis
forms an accurate, but non-orthogonal set. Hence the overlap matrix
\begin{equation}
\label{overlap}
S_{ij} = \int_0^\infty dr\,r\,B_i^{(k)}(r) B_j^{(k)}(r)
\end{equation}
is non-diagonal. It is diagonally banded, however, since $B_i^{(k)}(r)$
and $B_j^{(k)}(r)$ are functions that only overlap if $|i-j|\le k-1$.
The property of being diagonally banded also applies to the matrix
elements of other operators.

We now have a matrix equation with the $2\times 2$ block structure
\begin{equation}
\label{final}
\left\{\left[ 2 g_v V^+ - i\textstyle{\partial\over \partial z}\right]
\otimes I + {\cal H} {1\over \left[-i\textstyle{\partial\over \partial
z}\right]\otimes I} {\cal H} \right\} \left(\begin{array}{l}
\alpha^+\\\beta^+\end{array} \right) = p^- [1]\otimes I
\left(\begin{array}{l} \alpha^+\\\beta^+\end{array} \right),
\end{equation}
where $\otimes$ denotes an outer product of matrices. $\alpha^+ =
\{\alpha_{n,i}^+\}$ and $\beta^+ = \{\beta_{n,i}^+\}$ are column
vectors of length $(N\times {\cal N})$, and the $(N\times{\cal
N})\times (N\times {\cal N})$ sub-blocks have matrix elements
\begin{eqnarray}
\left[V^+\right]_{(nn'),(ij)} &=& \int_0^\infty
dr\,r\,B_i^{(k)}(r) B_j^{(k)}(r) \left[V^+(r)\right]_{(nn')}\, ,\\
\left[M^*\right]_{(nn'),(ij)} &=& \int_0^\infty
dr\,r\,B_i^{(k)}(r) B_j^{(k)}(r) \left[M^*(r)\right]_{(nn')}\, ,\\
\left[-i \textstyle{\partial \over \partial
z}\right]_{(nn'),(ij)} &=& p_n^+
\delta_{n,n'} S_{ij}\, ,\\
\left[1\right]_{(nn'),(ij)} &=& \delta_{n,n'} S_{ij}\, ,\\
\left[D_1\right]_{(nn'),(ij)} &=& \delta_{n,n'}\,\int_0^\infty
dr\,r\,B_i^{(k)}(r) \left( - {d\over dr}B_j^{(k)}(r)-{j_z+\case1/2\over r}
B_j^{(k)}(r)\right),\\
\left[D_2\right]_{(nn'),(ij)} &=& \delta_{n,n'}\,\int_0^\infty
dr\,r\,B_i^{(k)}(r) \left( {d\over dr}B_j^{(k)}(r)-{j_z-\case1/2\over r}
B_j^{(k)}(r)\right),\\
\left[i {\partial \Lambda\over \partial r}\right]_{(nn'),(ij)} &=&
- \int_0^\infty
dr\,{d\over dr}\left( r\,B_i^{(k)}(r) B_j^{(k)}(r) \right)
\left[i \Lambda(r)\right]_{(nn')}\, .
\end{eqnarray}
The last relation follows from using integration by parts.

\subsection{Numerical methods}

For our numerical calculations we use $L=12$~fm and $N=24$ Fourier
components, and set the number of terms in the cosine expansion of the
densities to be $N/2$, or 12. We choose $k=5$ for the degree of the
B-splines, ${\cal N} = 20$ B-splines in the expansion in $r$, and take
$0<r<L$. The integrals over $r$ are performed using Gaussian
integration between knots, which gives exact results for the matrix
elements $S_{ij}$, Eq.~(\ref{overlap}).

The matrix eigenvalue problem Eq.~(\ref{final}) is of the form $A x =
\lambda B x$, where $A$ and $B$ are real, symmetric matrices. In our
problem, $A$ and $B$ are diagonally banded, and there are efficient
EISPACK routines that take advantage of this \cite{eispack}. Cholesky
decomposition is used to efficiently compute the matrix
\begin{equation} {1\over \left[ -i {\partial\over \partial
z}\right]\otimes I} {\cal H}, \end{equation} which is needed both to
determine $u^-$ and $l^-$ as well as to construct the matrix in
Eq.~(\ref{final}).

\section{Momentum Distributions}
\label{appb}

In momentum space the meson field equations become
\begin{eqnarray}
\phi(k^+,\bbox{k}_\perp) &=& -{g_s\over \bbox{k}^2 + m_s^2}
\rho_s(k^+,\bbox{k}_\perp),\\
V^+(k^+,\bbox{k}_\perp) &=& {g_v\over \bbox{k}^2 + m_s^2}
\rho^+(k^+,\bbox{k}_\perp),
\end{eqnarray}
with $\bbox{k}^2 = (k^+)^2 + \bbox{k}_\perp^2$, and the convention
\begin{equation}
V^+(k^+,\bbox{k}_\perp) = \int d^2 x_\perp\,\int dz\,
e^{-i \bbox{k}_\perp\cdot \bbox{x}_\perp - i k^+ z}
V^+(z,\bbox{x}_\perp).
\end{equation}

The scalar meson momentum distribution from Eq.~(\ref{f1phi}) can be
rewritten as
\begin{eqnarray}
f_s(k^+) &=& {2 k^+ \over 2 \pi} \int {d^2 k_\perp \over (2\pi)^2}\,
{g_s^2\over (\bbox{k}^2 + m_s^2)^2} |\rho_s(k^+,\bbox{k}_\perp)|^2\\
&=& {2 k^+ \over 2 \pi} \int {d^2 k_\perp \over (2\pi)^2}\,
|\phi(k^+,\bbox{k}_\perp)|^2\\
&=& {2 k^+ \over 2 \pi} \int d^2 x_\perp \, |\phi(k^+,\bbox{x}_\perp)|^2.
\end{eqnarray}
Then
\begin{eqnarray}
P_s^+ &=& \int_0^\infty dk^+\, k^+ f_s(k^+)\\
&=& {1\over 2} \int_{-\infty}^\infty {dk^+\over 2\pi}\, 2 (k^+)^2\,
\int d^2 x_\perp\,|\phi(k^+,\bbox{x}_\perp)|^2\\
&=& \int d^3 x\, (\partial^+ \phi(\bbox{x}))^2\\
&=& \langle T_s^{++} \rangle.
\end{eqnarray}

The vector meson momentum distribution is a little more complicated. 
Starting from Eq.~(\ref{f2v}),
\begin{eqnarray}
f_v(k^+) &=& {2 k^+\over 2\pi} \int {d^2 k_\perp \over (2\pi)^2}\,
{g_v^2\over (\bbox{k}^2 + m_v^2)^2} {k_\perp^2 + m_v^2\over (k^+)^2}
|\rho^+(k^+,\bbox{k}_\perp)|^2\\
&=& {2 k^+ \over 2 \pi} \int {d^2 k_\perp \over (2\pi)^2}\,
{g_v\over \bbox{k}^2 + m_v^2} {\bbox{k}^2 + m_v^2-(k^+)^2\over (k^+)^2}
V^+(k^+,\bbox{k}_\perp) \rho^+(k^+,\bbox{k}_\perp)\\
&=& {2\over 2\pi} \int {d^2 k_\perp \over (2\pi)^2}\,
\left[{1\over k^+} g_v V^+(k^+,\bbox{k}_\perp) \rho^+(k^+,\bbox{k}_\perp) -
k^+ |V^+(k^+,\bbox{k}_\perp)|^2\right].
\end{eqnarray}
Then
\begin{eqnarray}
P_v^+ &=& \int_0^\infty dk^+\, k^+ f_v(k^+)\\
&=& {1\over 2} \int_{-\infty}^\infty {dk^+\over 2\pi}\, 2 \int d^2 x_\perp\,
\left[g_v V^+(k^+,\bbox{x}_\perp) \rho^+(k^+,\bbox{x}_\perp)-
(k^+)^2 |V^+(k^+,\bbox{x}_\perp)|^2 \right]\\
&=& \int d^3 x\, \left[ g_v V^+(\bbox{x}) \rho^+(\bbox{x}) -
(\partial^+ V^+(\bbox{x}))^2\right]\\
&=& \langle T_v^{++} \rangle.
\end{eqnarray}
Clearly $f_v(k^+)$ is singular at $k^+=0$, so we plot $k^+ f_v(k^+)$
instead.

Momentum distributions involve integrals over $\bbox{x}_\perp$, or
equivalently over $\bbox{k}_\perp$, so one really only needs to Fourier
transform $V^+(z,\bbox{x}_\perp)$ in $z$. If we define
\begin{equation}
V^+(k^+,\bbox{x}_\perp) = \int dz\, e^{-i k^+ z} V^+(z,\bbox{x}_\perp),
\end{equation}
then for $k_m^+ = m q, q=\pi/L$, it follows from the definition
Eq.~(\ref{potexpand}) that
\begin{equation}
V^+(k_m^+,\bbox{x}_\perp) = 2 L {1+\delta_{m,0}\over 2} V_m^+(r),
\end{equation}
with a similar result for $\rho^+(k_m^+,\bbox{x}_\perp)$,
$\phi(k_m^+,\bbox{x}_\perp)$, and $\rho_s(k_m^+,\bbox{x}_\perp)$. Hence
we calculate the momentum distributions from the expressions
\begin{eqnarray}
k_m^+ f_s(k_m^+) &=& {2(m q)^2 \over q} {(1+\delta_{m,0})^2\over 4}
2\pi\,2L \int_0^\infty dr\,r (\phi_m(r))^2\\
k_m^+ f_v(k_m^+) &=& {2 \over q}
{(1+\delta_{m,0})^2\over 4} 2\pi\,2L \int_0^\infty dr\,r \left[
g_v V_m^+(r) \phi_m^+(r) - (m q)^2 (V_m^+(r))^2 \right].
\end{eqnarray}

The nucleon momentum distribution for $p_n^+ = n q$ is given by
\begin{equation}
f_N(p_n^+) = {2\over q} \int_0^\infty dr\, r \sum_\alpha^{\text{occ}}
\left[(u_{\alpha,n}^+(r))^2 + (l_{\alpha,n}^+(r))^2\right],
\end{equation}
so that $A = \int_0^\infty dp^+\, f_N(p_n^+) \approx q \sum_n
f_N(p_n^+)$, and $P_N^+ = \int_0^\infty dp^+\, p^+ f_N(p_n^+) \approx q
\sum_n p_n^+ f_N(p_n^+)$. We interpolate between the discrete values of
$k_n^+$ and $p_n^+$ to produce the plots of Figs.~1 and 2.


\begin{references}

\bibitem{emc} J. Aubert {\it et al.}, Phys.\ Lett.\ 
{\bf 123B}, 275 (1982);
R.G. Arnold {\it et al.,} Phys. Rev. Lett. {\bf 52}, 727 (1984);
A. Bodek {\it et al.}, Phys. Rev. Lett. {\bf 51}, 534 (1983).

\bibitem{fs} L.L.~Frankfurt and M.I.~Strikman,
Phys. Rep. {\bf 76}, 215 (1981).

\bibitem{mb:adv} M. Burkardt, Adv. Nucl. Phys. {\bf 23}, 1 (1996).

\bibitem{emcrevs}R.L. Jaffe,
in {\it Relativistic Dynamics and Quark-Nuclear Physics},
edited by M.B. Johnson and A. Picklesimer (Wiley, New York, 1985); 
L.L. Frankfurt and M.I. Strikman, Phys. Rep. {\bf 160}, 235 (1988);
M. Arneodo, Phys. Rep. {\bf 240}, 301 (1994);
D.F. Geesaman,
K. Saito, A.W. Thomas, Ann. Rev. Nucl. Part. Sci. {\bf 45}, 337 (1995).

\bibitem{west} G.B. West, Phys. Rep. {\bf 18C}, 264 (1975). 

\bibitem{jf} X. Ji and B. W. Filippone, Phys. Rev. C {\bf 42}, R2279
(1990).

\bibitem{sdm} I. Sick, D. Day, and J.S. McCarthy, Phys. Rev. Lett. {\bf 45},
871 (1980).

\bibitem{dyth}R.P. Bickerstaff, M.C. Birse, and G.A. Miller,
Phys. Rev. Lett. {\bf 53}, 2532 (1984); M. Ericson and A.W. Thomas,
Phys. Lett. {\bf 148B}, 191 (1984); E.L. Berger, Nucl. Phys. {\bf B267},
231 (1986).
 
\bibitem{dyexp} D.M. Alde {\it et al.}, Phys. Rev. Lett. {\bf 64}, 2479 (1990).

\bibitem{missing} G.F. Bertsch, L. Frankfurt, and M. Strikman,
Science {\bf 259}, 773 (1993).

\bibitem{jerry} G.A. Miller, Phys. Rev. C {\bf 56}, R8 (1997);
{\bf 56}, 2789 (1997).

\bibitem{bsjdw}B.D. Serot and J.D. Walecka, Adv. Nucl. Phys. {\bf 16},
1 (1986);
Int. J. Mod. Phys. {\bf E6}, 515 (1997).

\bibitem{harizhang}
A. Harindranath and W.-M. Zhang, Phys. Rev. D {\bf 48}, 4861 (1993);
{\bf 48}, 4881 (1993); {\bf 48}, 4903 (1993).

\bibitem{us} P.G.~Blunden, M.~Burkardt, and G.A.~Miller, Phys. Rev. C (to
be published), nucl-th/9901063.

\bibitem{us2}
P.G.~Blunden, M.~Burkardt and G.A.~Miller, {\it Light-Front Nuclear
Physics: Toy Models, Static Sources and Tilted Light-Front
Coordinates}, preprint NT@UW-99-22, to be submitted to Phys. Rev C.
 
\bibitem{notation} Our notation is that a four vector $A^\mu$ is defined by
the plus, minus and perpendicular components as
$(A^0+A^3,A^0-A^3,\bbox{A}_\perp)$ and $A\cdot B={1\over 2} A^+B^-+{1\over
2}A^-B^+-\bbox{A_\perp\cdot B_\perp}$.

\bibitem{lcrevs} S. J. Brodsky, H-C Pauli,
S.S.~Pinsky, Phys. Rep. {\bf 301}, 299 (1998);
S.J.~Brodsky and G.P.~Lepage,
in {\it Perturbative Quantum Chromodynamics}, edited by A. Mueller,
(World Scientific, Singapore, 1989); X.D. Ji,
Comm. Nucl. Part. Phys. {\bf 21}, 123 (1992);
W.-M. Zhang,
Chinese J. Phys. {\bf 32}, 717 (1994);
{\it Theory of hadrons and light-front QCD}, edited by S.D. Glazek, (World
Scientific, Singapore, 1994).

\bibitem{des71}D.E. Soper, {\it Field Theories in the Infinite Momentum
Frame},
SLAC pub-137 (1971); 
D.E. Soper, Phys. Rev. D {\bf 4}, 1620 (1971); J.B. Kogut and D.E. Soper,
Phys. Rev. D {\bf 1}, 2901 (1971).

\bibitem{yan12} S.-J. Chang, R.G. Root, and T.-M. Yan, Phys. Rev. D
{\bf 7}, 1133 (1973); {\bf 7}, 1147 (1973).

\bibitem{yan34}T.-M. Yan, Phys. Rev. D {\bf 7}, 1760 (1974); {\bf
7}, 1780 (1974).

\bibitem{mpsw} D.~Mustaki, S.~Pinsky, J.~Shigemitsu, and K.~Wilson,
Phys. Rev. {\bf D43}, 3411 (1991).

\bibitem{bm98} M.~Burkardt and G.A.~Miller,
Phys. Rev. C {\bf 58}, 2450 (1998).

\bibitem{cc} Consider the term $ i\partial_\mu\bar
\psi\gamma^\mu\psi=0= {i\over2}\partial^+\bar \psi\gamma^-\psi
+{i\over2}\partial^-\bar \psi\gamma^+\psi-
i\bbox{\nabla}_\perp\cdot\bar \psi \bbox{\gamma}_\perp \psi.$ The term
with $\partial^-$ vanishes, because the fermion scalar density is
independent of $x^+$. Then taking the Fourier transform as needed in
Eq.~(\ref{delta}) leads to the equality (\ref{equal}) for the case
$\mu=-$.

\bibitem{hs} C.J. Horowitz and B.D. Serot, Nucl. Phys. {\bf A368}, 503 (1981).
We neglect here the electromagnetic effects as well as those of the
$\rho$ meson.

\bibitem{frank} H. Kim, C.J. Horowitz, and M.R. Frank,
Phys.Rev. C {\bf 51}, 792 (1995).

\bibitem{ls} R.J. Furnstahl, J.J. Rusnak, and B.D. Serot,
Nucl. Phys. {\bf A632}, 607 (1998).

\bibitem{rmgm98}G.A. Miller and R. Machleidt, Phys. Lett. B (to be
published), nucl-th/9811050;
G.A.~Miller and R.~Machleidt, Phys. Rev. C. to be published.
{\it Infinite nuclear matter on the light-front: Nucleon-nucleon
correlations}, nucl-th/9903080.

\bibitem{zim} J.~Zimanyi, S.A.~Moszkowski, Phys. Rev. C {\bf 42}, 1416 (1990);
N.K.~Glendenning, F. Weber, and S.A.~Moszkowski,
Phys. Rev. C {\bf 45}, 844 (1992); S.A.~Moszkowski,
(private communication).

\bibitem{qmc}
K.~Saito and A.W.~Thomas,
Phys. Lett. {\bf 327B}, 9 (1994);
P.G.~Blunden and G.A.~Miller,
Phys. Rev. C {\bf 54}, 359 (1996);
X.~Jin and B.K.~Jennings,
Phys. Rev. C {\bf 54}, 1427 (1996).

\bibitem{deboor} C. deBoor, {\it A Practical Guide to Splines}
(Springer, New York, 1978).

\bibitem{mcneil} J.A. McNeil {\it et al.}, Phys. Rev. C {\bf 40}, 399
(1989); W. Johnson, S. Blundell, and J. Sapirstein, Phys. Rev. A {\bf
37}, 307 (1988).

\bibitem{eispack} E. Anderson {\it et al.}, {\it LAPACK Users Guide}
(SIAM, Philadelphia, 1994), 2nd Ed.

\end{references}
\end{document}